\documentclass[iop, revtex4]{emulateapj}

\usepackage{hyperref, graphicx, natbib, multirow, captcont, amsmath}

\newcommand{\um}{$\mu$m}
\newcommand{\latemovers}{46,463}
\DeclareRobustCommand{\rchi}{{\mathpalette\irchi\relax}}
\newcommand{\irchi}[2]{\raisebox{\depth}{$#1\chi$}} 

\slugcomment{}
\shorttitle{LaTE-MoVeRS}
\shortauthors{Theissen et al.}

\begin{document}

\title{The Late-Type Extension to M\MakeLowercase{o}V\MakeLowercase{e}RS (L\MakeLowercase{a}TE-M\MakeLowercase{o}V\MakeLowercase{e}RS): Proper motion verified low-mass stars and brown dwarfs from SDSS, 2MASS, and \emph{WISE}}

\author{Christopher A. Theissen\altaffilmark{1}, Andrew A. West}
\affil{Department of Astronomy, Boston University, Boston, MA 02215, USA}
\altaffiltext{1}{Visiting Graduate Student, University of California, San Diego}
\email{ctheisse@bu.edu}
\author{Guillaume Shippee}
\affil{Department of Physics, University of California, Berkeley, CA 94720, USA}
\author{Adam J. Burgasser}
\affil{Department of Physics, University of California, San Diego, CA 92093, USA}
\and
\author{Sarah J. Schmidt}
\affil{Leibniz-Institute for Astrophysics Potsdam (AIP), An der Sternwarte 16, D-14482, Potsdam, Germany}

\begin{abstract}

We present the Late-Type Extension to the Motion Verified Red Stars (LaTE-MoVeRS) catalog, containing \latemovers\ photometric late-type ($>$ M5) dwarfs within the Sloan Digital Sky Survey (SDSS) footprint. 
Proper motions were computed for objects combining astrometry from the SDSS Data Release 12 (DR12), the Two-Micron All-Sky Survey (2MASS) Point Source Catalog (PSC), and the \emph{Wide-field Infrared Survey Explorer} (\emph{WISE}) AllWISE datasets. 
LaTE-MoVeRS objects were required to have significant proper motion ($\mu_\mathrm{tot} \geqslant 2\sigma_{\mu_\mathrm{tot}}$). 
Using the LaTE-MoVeRS sample and \emph{Gaia} Data Release 1, we estimate \emph{Gaia} will be $\sim$64\% complete for very-low-mass objects ($>$ M5) in comparison to the combined SDSS+2MASS+\emph{WISE} dataset ($i < 21.3$).
We computed photometric distances and estimated stellar effective temperatures for the LaTE-MoVeRS catalog. 
The majority of the dwarfs in the sample have distances $< 150$ pc and $T_\mathrm{eff} < 3000$ K. 
Thirteen objects were identified within LaTE-MoVeRS with estimated photometric distances within 25 pc that have not been previously identified as nearby objects. 
We also identified one new object with a large amount of excess mid-infrared flux that has not been previously identified (2MASS J11151597$+$1937266). 
This object appears to be an L2$\gamma$ at $\sim$50 pc showing spectroscopic signs of a flaring event (e.g., strong Hydrogen Balmer emission lines). 
This object does not exhibit kinematics similar to any known kinematic association.
The LaTE-MoVeRS catalog is available through SDSS CasJobs and VizieR.

\end{abstract}


\section{Introduction}\label{intro}
 
	Very-low-mass (VLM) stars and brown dwarfs \citep[spectral types $>$ M5, $M_\ast < 0.15M_\odot$;][]{baraffe:1996:l51} are ubiquitous and among the longest living objects in the Galaxy. This makes these objects extremely useful for studies of Galactic kinematics, the luminosity and mass functions, and exoplanet searches \citep[e.g.,][]{luhman:2000:1016, reid:2002:2721, bochanski:2010:2679, berta:2012:145, gillon:2016:221, van-vledder:2016:425}. Additionally, these objects are important for testing different formation scenarios for VLM objects \citep[e.g.,][]{reipurth:2001:432, padoan:2002:870, whitworth:2004:299, stamatellos:2009:413}. Combining photometric VLM objects with proper motions further allows for the discovery and analysis of common proper motion binaries, which can test formation mechanisms and evolutionary models \citep{jameson:2008:1399}.
 
	A number of programs are underway to obtain a complete census of nearby ($<$30 pc) stars and brown dwarfs, such as the REsearch Consortium On Nearby Stars \citep[RECONS;][]{riedel:2014:85, winters:2015:5} and the SUPERBLINK proper motion survey \citep{lepine:2002:1190, lepine:2003:921, lepine:2005:1483}. However, these censuses are estimated to be incomplete by as much as $\sim$30\% \citep[][]{winters:2015:5}. These studies use significant proper motions to identify objects that have a high-probability of being nearby, low-mass objects versus more distant giants and extragalactic objects with similar colors. To identify the missing members of these nearby samples---slow-moving objects \citep[e.g., \emph{WISE} J072003.20-084651.2;][]{scholz:2014:a113} and/or young objects---higher-precision proper motion catalogs must be developed.
 
	Current, high-precision, proper motion catalogs, such as the United States Naval Observatory (USNO) B1.0 catalog \citep[USNO-B1.0;][]{monet:2003:984} and the USNO CCD Astrograph Catalog \citep[UCAC;][]{zacharias:2000:2131,zacharias:2004:3043,zacharias:2010:2184,zacharias:2013:44}, are biased towards objects bluer than VLM objects \citep{theissen:2016:41}. The currently underway \emph{Gaia} mission will achieve an unprecedented astrometric precision \citep[typical end-of-mission parallax errors $\lesssim$16 $\mu$as for an M6 with $V \lesssim 14$;][]{perryman:2001:339}. However, \emph{Gaia} is estimated to be incomplete for VLM objects relative to surveys such as the Sloan Digital Sky Survey \citep[SDSS;][]{york:2000:1579}, largely due to the magnitude limit of the astrometric instrument \citep[$r \lesssim 20$ versus $r \leqslant 22.2$;][]{ivezic:2008:}. Additionally, \emph{Gaia} operates at visible wavelengths, making it limited in studies of very cool (red) objects. Conversely, mid-infrared proper motion surveys optimized to find very cool objects, such as AllWISE \citep{kirkpatrick:2014:122}, have poor precision ($\sigma_{\mu_\mathrm{tot}} \approx 100$ mas yr$^{-1}$), and are only able to measure objects with relatively large proper motions ($\gtrsim 150$ mas yr$^{-1}$). This motivates the need for higher-precision, proper motion catalogs optimized for VLM objects, specifically measuring proper motions in the near- and mid-infrared (NIR/MIR), where the spectral energy distributions (SEDs) of these VLM objects peak.

	One of the largest searches for VLM objects was undertaken by \citet{gagne:2015:73}, compiling the Bayesian Analysis for Nearby Young AssociatioNs (BANYAN) All-Sky Survey (BASS) catalog. By cross-matching sources from the Two-Micron All-Sky Survey \citep[2MASS;][]{skrutskie:2006:1163} and the \emph{Wide-field Infrared Survey Explorer} \citep[\emph{WISE};][]{wright:2010:1868}, \citet{gagne:2015:73} produced a catalog of 98,970 VLM objects with expected spectral types later than M4. One limitation of the BASS catalog is the requirement that objects have $J-K_s > 0.77$, which was put in place to specifically isolate young VLM stars and brown dwarfs, excluding many field objects. Additionally, the requirement that objects have a minimum total proper motion greater than 40 mas yr$^{-1}$ may also exclude many nearby objects (e.g., \emph{WISE} J072003.20-084651.2) according to galactic models \citep{theissen:2016:2}. This leaves a relatively large color space open to investigation, and highlights the need for higher precision proper motion measurements for the reddest and faintest objects.
	
	Currently, the largest proper motion catalog optimized for low-mass stars is the Motion Verified Red Stars catalog \citep[MoVeRS;][]{theissen:2016:41}. MoVeRS contains 8,735,004 stars, and was built using photometric and astrometric data from the Sloan Digital Sky Survey \citep[SDSS;][]{york:2000:1579}, 2MASS, and \emph{WISE}. However, the requirement that all objects within MoVeRS have SDSS $r$-band detections makes it incomplete for objects with spectral-types later than $\sim$M7 (past a distance of $\sim$100 pc). Most studies of VLM objects from SDSS exclude $r$-band measurements, including only $i$- and $z$-band detections \citep[e.g.,][]{schmidt:2010:1808,schmidt:2015:158}. To address the red nature of late-type dwarfs, an extension of MoVeRS can be constructed using $i$-band astrometry from SDSS in conjunction with 2MASS and \emph{WISE} astrometry.
	
	This paper outlines the methods used in constructing the Late-Type Extension to MoVeRS (LaTE-MoVeRS). The LaTE-MoVeRS sample consists of \latemovers\ objects with $i-z > 1.04$ (spectral-types later than $\sim$M5). Section~\ref{methods} outlines the methods used in reconstructing SDSS $i$-band astrometry, and computing proper motions with the combined SDSS, 2MASS, and \emph{WISE} datasets. The criteria used in selecting sources for the LaTE-MoVeRS, and testing the fidelity of the LaTE-MoVeRS proper motions are described in Section~\ref{data}. We estimate the incompleteness of \emph{Gaia} for VLM stars and brown dwarfs using Data Release 1 and our LaTE-MoVeRS sample in Section~\ref{Gaia}. The general characteristics, including color, distance, and temperature distributions of the LaTE-MoVeRS catalog are discussed in Section~\ref{latemovers}. We also discuss new, nearby systems in Section~\ref{distancetemp}. In Section~\ref{mirexcess} we use the LaTE-MoVeRS sample to search for stars exhibiting excess MIR flux. Lastly, in Section~\ref{discussion} we provide a discussion and summary of our work.

\section{Methods: Combining \emph{WISE}, SDSS, and 2MASS}\label{methods}

\subsection{Astrometric Algorithms}\label{pmalg}

	To measure robust proper motions, the astrometric positions and uncertainties for each of the surveys (epochs) used must be fully understood. The surveys used for this study are SDSS, 2MASS, and \emph{WISE}, which provide a wavelength range of $\sim$7500--35000\AA, sampling the peak of the spectral energy distributions (SEDs) for late-types dwarfs. We will give a brief description of the astrometric methods used in this study; a more detailed explanation can be found in \citet{theissen:2016:41}. In this section, we focus on the methods used to reconstruct the SDSS $i$-band astrometry. 
	
	The SDSS absolute astrometry is measured in the $r$-band, calibrated directly to UCAC \citep{zacharias:2000:2131} and Tycho-2 \citep{pier:2003:1559}. However, for the reddest objects, the $r$-band photometry suffers from relatively low counts, making these objects absent from both the UCAC and Tycho-2 catalogs. For objects detected in the other SDSS bands, a secondary astrometric catalog is made using SDSS astrometrically calibrated stars from Tycho-2 and UCAC. This means a reconstruction of astrometry in the other bands is possible, with slightly higher uncertainties than the $r$-band astrometry, due to the systematic offsets between the band-to-band astrometric transformations. Here, we describe the methods used in recreating $i$-band astrometry.

	To correct for optical distortions and differential chromatic refraction (DCR), we used the following equations (taken from \citealt{pier:2003:1559}):
\begin{equation}
x^\prime = x + g_0 + g_1 y + g_2 y^2 + g_3 y^3 + p_x (color),
\end{equation}
and
\begin{equation}
y^\prime = y + h_0 + h_1 y + h_2 y^2 + h_3 y^3 + p_y (color),
\end{equation}
where $x$ is the frame row pixel (\textsc{rowc}\_i), $y$ is the frame column pixel (\textsc{colc}\_i), $g_n$ are the row distortion coefficients \textsc{dRow}$n$, $h_n$ are the column distortion coefficients \textsc{dCol}$n$, and $p_x$ and $p_y$ are the DCR corrections \textsc{ccRow}\_i and \textsc{ccCol}\_i, respectively. These values are found in the CasJobs\footnote{\url{http://skyserver.sdss.org/casjobs/}} \textsc{Field} and \textsc{PhotoObjAll} tables. 

\begin{figure}
\centering
 \includegraphics[width=\linewidth]{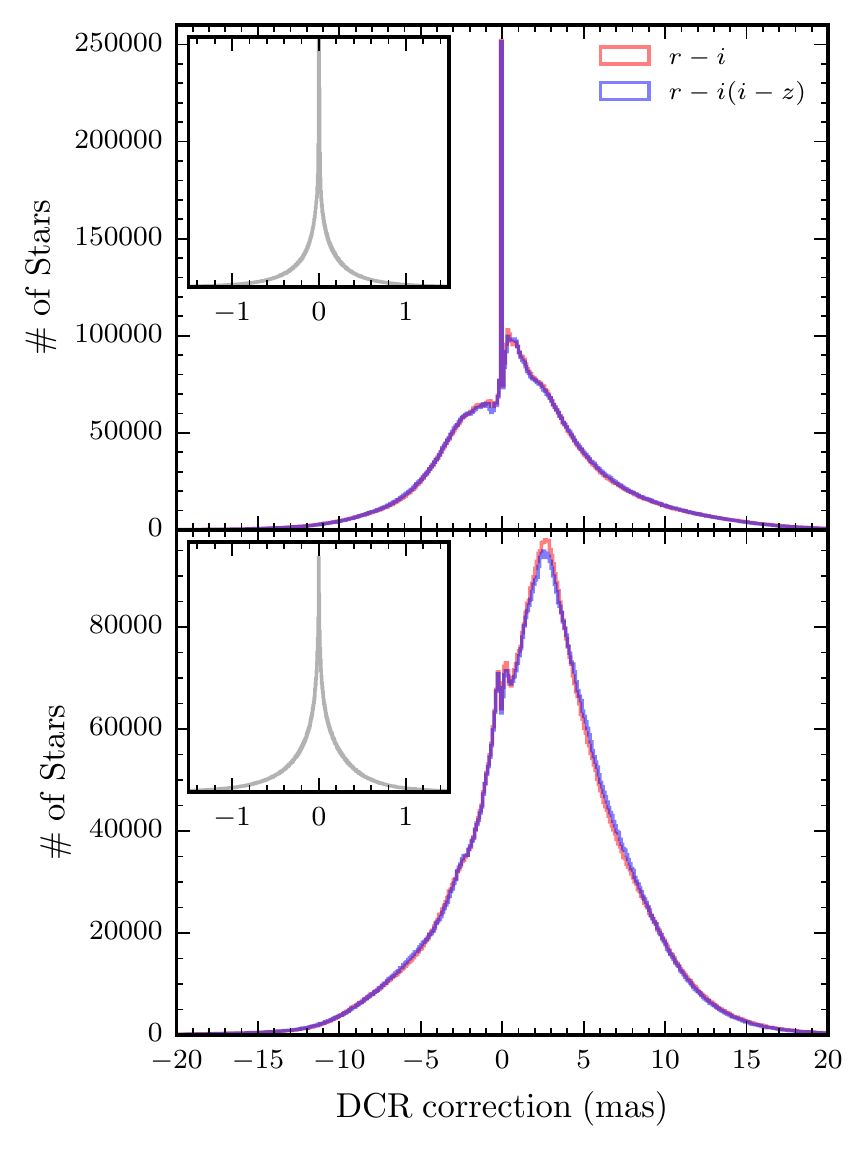}
\caption{DCR corrections using both actual $r-i$ color (red) and $r-i$ color from Equation~(\ref{eqn:riiz}) (blue). Stars are from the MoVeRS sample. The corrections are well approximated using $i-z$ color as a proxy for $r-i$ color. The top plot shows right ascension and the bottom plot shows declination. Inset plots show the residuals between the two distributions. Although the residuals are non-Gaussian, the uncertainty added by this transformation is negligent ($< 0.4$ mas; see text).
\label{fig:DCR}}
\end{figure}

	The DCR correction is computed as a linear equation of $r-i$ color. However, for many of the late-type dwarfs, the $r$-band photometry is at or below the magnitude limit of SDSS, and is therefore unreliable. It is far more preferable to use $i-z$ color for our DCR corrections. $r-i$ and $i-z$ color are strongly correlated for low-mass stars \citep[see Figure 4;][]{theissen:2016:41}. To remove the need for $r-i$ color, we performed a linear fit to the $r-i$ versus $i-z$ color-color diagram using stars with colors between $0.2 < i-z < 1.4$ from the MoVeRS sample. Our fit is given by,
\begin{equation}\label{eqn:riiz}
r-i = 1.797(i-z) - 0.008\quad (0.2 < i-z < 1.4).
\end{equation} 
This equation yields an estimate of $r-i$ color for the DCR correction, with the calculated error in the color residual, $(r-i) - $Equation~(\ref{eqn:riiz}), of $\sigma \approx 0.08$ mag. For objects redder than $i-z =1.4$, we used the \citet{schmidt:2010:1808} spectroscopic sample of 484 L dwarfs from SDSS. The $r-i$ color is approximately constant with $i-z$ color, due to the similar slopes in the Wein tails for these red objects. Comparing to stars with $r$-band magnitudes $\leqslant 22.2$ (184 dwarfs; the SDSS 95\% completeness limit for the $r$-band), we found the best constant value was given by,
\begin{equation}\label{eqn:riiz2}
r-i = 2.5\quad (i-z \geqslant 1.4).
\end{equation}
The fit uncertainties from Equations (\ref{eqn:riiz}) and (\ref{eqn:riiz2}) introduce typical uncertainties on the order of 0.01 pixels, and are therefore negligible. The DCR corrections using actual $r-i$ colors and $r-i$ colors from Equation~(\ref{eqn:riiz}) are shown in Figure~\ref{fig:DCR}. From the residuals between actual and approximated DCR corrections, we quantified the uncertainty added by this procedure by calculating the 16$^\mathrm{th}$ and 84$^\mathrm{th}$ percentile values for the residual distributions. These values were $-$0.29 mas and 0.29 mas for the right ascension, respectively, and -0.34 mas and 0.33 mas for the declination, respectively. These values are much smaller than the SDSS astrometric uncertainty which is $\sim$40 mas\footnote{\url{http://classic.sdss.org/dr7/algorithms/astrometry.html}}, and hence negligible.

The corrected pixel coordinates can then be transformed into catalog mean place great circle longitude ($\mu$) and latitude ($\nu$) using the equations:
\begin{equation}
\mu = a_i + b_i x^\prime + c_i y^\prime, 
\end{equation}
and
\begin{equation}
\nu = d_i + e_i x^\prime + f_i y^\prime,
\end{equation}
where $a_i$, $b_i$, $c_i$, $d_i$, $e_i$, $f_i$ are the astrometric coefficients in the \textsc{Field} table for the $i$-band. 

	Next, these great circle coordinates must be transformed into J2000.0 $\alpha$ and $\delta$ coordinates in degrees using the equations
\begin{equation}
\alpha = \tan^{-1} \left[ \dfrac{\sin(\mu - \alpha_0)\cos \nu \cos i - \sin \nu \sin i}{\cos(\mu - \alpha_0) \cos \nu} \right] + \alpha_0,
\end{equation}
and
\begin{equation}
\delta = \sin^{-1} [ \sin(\mu - \alpha_0)\cos \nu \sin i - \sin \nu \cos i ],
\end{equation}
where $\alpha_0$ and $i$ are the right ascension and inclination of the ascending node of the great circle with respect to the J2000.0 celestial equator, respectively, both of which are found in the \textsc{Frame} table. Errors in $\alpha$ and $\delta$ can be computed using the methods described in \citet{theissen:2016:41} using the $i$-band coefficients.

	We use the catalog astrometry for 2MASS, which is tied to the International Celestial Reference System (ICRS) through Tycho-2 \citep{hog:2000:l27}.	2MASS astrometric errors are reported as error ellipses, with the entries \textsc{err\_maj} ($\sigma_\mathrm{MAJ}$), \textsc{err\_min} ($\sigma_\mathrm{MIN}$), and \textsc{err\_ang} ($\sigma_\mathrm{\theta}$). These were converted to $\sigma_\alpha$ and $\sigma_\delta$ components using the following equations,
\begin{equation}
	\sigma_{\alpha_\mathrm{2MASS}} = \sqrt{(\sigma_\mathrm{MAJ} \cdot \sin\sigma_\theta)^2 + (\sigma_\mathrm{MIN} \cdot \cos\sigma_\theta)^2}
\end{equation}
and
\begin{equation}
	\sigma_{\delta_\mathrm{2MASS}} = \sqrt{(\sigma_\mathrm{MAJ} \cdot \cos\sigma_\theta)^2 + (\sigma_\mathrm{MIN} \cdot \sin\sigma_\theta)^2}.
\end{equation} 

	We also used catalog astrometry for \emph{WISE}, which is tied to the ICRS through 2MASS. \emph{WISE} astrometric errors are denoted by the entries \textsc{sigRA} ($\sigma_\textsc{sigRA}$) and \textsc{sigDEC} ($\sigma_\textsc{sigDEC}$). Further details regarding 2MASS and \emph{WISE} astrometry can be found in \citet{theissen:2016:41}.

\subsubsection{SDSS Astrometric Offsets}\label{offsets}

\begin{figure*}
\centering
\includegraphics[width=\linewidth]{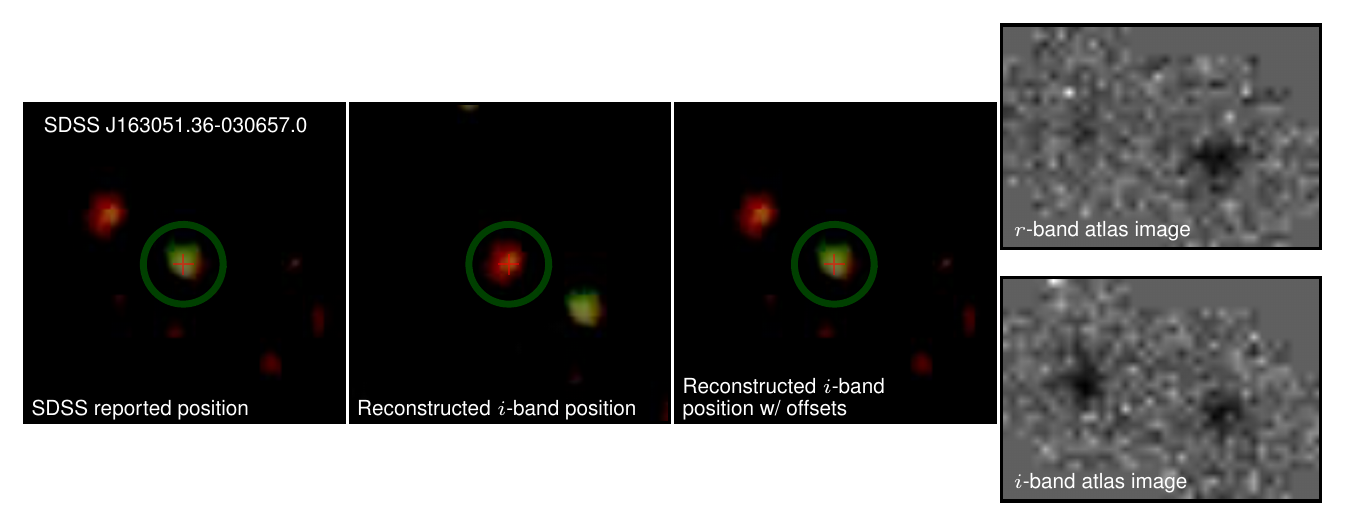}
\caption{$gri$-composite images (SDSS J163051.36-030657.0) demonstrating the blended object issue (Section~\ref{offsets}). Each image is $25\arcsec \times 25\arcsec$. \emph{Left}: Reported SDSS position with the green circle centered on the reported position. \emph{Middle Left}: Reconstructed $i$-band position using our DCR correction, again with the blue circle centered on the position. \emph{Middle Right}: Reconstructed $i$-band position using our DCR correction with the SDSS offsets applied, again with the blue circle centered on the position. The red object does not exist within the SDSS catalog because it is confused with the nearby object. \emph{Far Right}: SDSS Atlas images for the $r$- and $i$-band for this object. Each Atlas image should only reference a single object within SDSS, although two objects can be seen in both Atlas images. Removing objects with the \textsc{nodeblend} flag minimize these types of confused objects.
\label{fig:offsets}}
\end{figure*}

	During this study, we noticed that the offset terms (\textsc{offsetRa\_x} and \textsc{offsetDec\_x} in the \textsc{PhotoObjAll} table, where \textsc{x} refers to the specific SDSS band) applied to bands not considered the canonical band, sometimes confused two objects with each other. This was done if an object had $ugr$-band detections, but no $iz$-band detections, and was close to another object with $iz$-band detections, but with no $ugr$-band detections. An example of this is shown in Figure~\ref{fig:offsets}, where some of the photometry for one of the objects is applied to the other (SDSS J163051.36-030657.0). The photometry for each SDSS band is measured independently. So although these objects have confused photometry at the catalog level, the photometry for each band should be independent, and represent a single object within the SDSS Frame. For example, in Figure~\ref{fig:offsets}, the brightest object in the $i$-band Atlas image is not the brightest object in the $r$-band Atlas image. Therefore the $i$-band photometry for this object would be reported for the object on the left, and the the $r$-band photometry would be reported for the object on the right. Additionally, each SDSS Atlas image should correspond to a single object, however, two objects are clearly seen in each Atlas image in Figure~\ref{fig:offsets}. The majority of these objects are flagged as \textsc{nodeblend} in the SDSS CasJobs \textsc{PhotoObj} Table, because they really are two objects that were not separated. Removing objects with this flag minimizes contamination due to these types of confused objects. Additionally, since VLM objects are brightest in the $i$- and $z$-bands, and we did not use the bluer SDSS passbands, our photometry is unlikely to be confused as in the example above.
	
\section{Data}\label{data}

\subsection{Building the Catalog}\label{catalog}

	Numerous studies have made use of the VLM stars and ultracool dwarfs within SDSS \citep[e.g.,][]{hawley:2002:3409, west:2004:426}. \citet{schmidt:2010:1808} used 484 spectroscopic L dwarfs to study the color and kinematics of ultracool dwarfs. One of the largest samples of spectroscopic late-M and L dwarfs was compiled by \citet{schmidt:2015:158} using SDSS BOSS spectra, the BOSS Ultracool Dwarf (BUD) sample of 11,820 objects. To select objects that had SDSS colors typical of late-type stars and brown dwarfs, we applied selection criteria adapted from \citet{schmidt:2015:158}. The initial selection criteria applied to SDSS Data Release 12 \citep[DR12;][]{alam:2015:12} photometry were:
\begin{enumerate}
\item Point-source-like morphology (\textsc{Type} = 6). This removes most extragalactic and other extended sources.
\item No children from deblending (\textsc{nChild} = 0). This ensures a clean point-source-function (PSF) fit.
\item $i < 21.3$. This selects objects within the SDSS 95\% completeness limit for point-sources in the $i$-band\footnote{\url{http://www.sdss.org/dr12/scope/}}.
\item $z < 20.5$. This selects objects within the SDSS 95\% completeness limit for point-sources in the $z$-band$^4$.
\item $i-z > 1.04$. This selects objects with red colors expected of VLM stars and brown dwarfs. We chose a slightly bluer cutoff than \citet[$i-z > 1.14$;][]{schmidt:2015:158} to be more inclusive for slightly earlier-type objects. 
\end{enumerate}
To illustrate the typical spectral types included and volumes probed using the above selection criteria, we calculated median $i-z$ colors and standard deviations per optical spectral type using the \citet{west:2011:97} sample for spectral types $<$ M7 and \citet{schmidt:2015:158} sample for spectral types $>$ M5, shown in Table~\ref{tbl:colors}. The standard deviation of the color for each spectral type bin represents the intrinsic spread in color per spectral type rather than the uncertainty in the median color. We also calculated the approximate distance limits using the S. J. Schmidt et al. (2016, in preparation) $i-z$ photometric parallax relationship (Section~\ref{distancetemp}) corresponding to the SDSS 95\% completeness limits in Table~\ref{tbl:colors}.

We removed sources that had a flag in the $i$- or $z$-bands indicating: 
\begin{enumerate}
\item saturated photometry (\textsc{saturated}); 
\item more than 20\% of the flux was interpolated from the point spread function (PSF; \textsc{psf\_flux\_interpreted}), which indicates the flux measurement is suspect (e.g., interpolation over a cosmic ray or row of bad pixels); 
\item too few good pixels for an interpolated source, causing errors to be underestimated (\textsc{bad\_counts\_error}); 
\item object did not have a radial profile (\textsc{noprofile}), likely suffering from low counts making the measured photometry suspect; 
\item object included pixels that were not checked for peaks (\textsc{notchecked}), typical for saturated sources or sources close to the edges; 
\item the center of the object was in a not-checked region, typically close to the edges or for saturated sources (\textsc{notchecked\_center}); 
\item after deblending the object did not have a peak (\textsc{deblend\_nopeak}), potentially indicating a poor PSF fit or failed deblending, making the photometry and astrometry of this blended object suspect; and
\item object failed deblending (\textsc{nodeblend}), causing blended objects to remain blended.
\end{enumerate}
Additionally, we removed sources that had the following flags in the $i$-band:
\begin{enumerate}
\item centroiding failure caused center to be determined by peak pixel (\textsc{peakcenter}), which affects the astrometry of the $i$-band; and
\item center pixel was too close to interpolated pixel (\textsc{interp\_center}), also affecting the $i$-band astrometry.
\end{enumerate}

	These initial criteria produced 5,199,350 objects within the SDSS footprint ($\sim$1/3 of the sky in the northern hemisphere). The photometric criteria described above go to, but do not exceed, the 95\% estimated completeness limits of SDSS point-sources, and should reduce the number of point-source-like extragalactic objects in our sample. The flags we selected on (listed above) help to ensure that we selected a nearby galactic sample as opposed to more distant objects. For this sample, we computed $i$-band astrometric positions following the methods outlined in Section~\ref{pmalg}.

\begin{deluxetable}{cccc}
\tabletypesize{\footnotesize}
\tablecolumns{4}
\tablecaption{Colors and Distances of VLM Dwarfs\label{tbl:colors}}
\tablehead{
\colhead{Spectral} & \colhead{Median $i-z$} & \colhead{Distance Limit\tablenotemark{b}} & \colhead{Distance Limit\tablenotemark{c}} \\
\colhead{type} & \colhead{color\tablenotemark{a}} & \colhead{(pc)} & \colhead{(pc)} 
}
\startdata
M5 	& 0.86 (0.11) 	& 615 	& 702	\\
M6 	& 1.06 (0.07)	& 428 	& 531	\\
M7 	& 1.17 (0.11) 	& 242 	& 337	\\
M8 	& 1.48 (0.14) 	& 186 	& 271	\\
M9 	& 1.66 (0.14) 	& 123 	& 194	\\
L0 	& 1.85 (0.10) 	& 105 	& 171	\\
L1 	& 1.86 (0.10) 	& 94 		& 153	\\
L2 	& 1.85 (0.08) 	& 89 		& 145	\\
L3 	& 1.88 (0.13) 	& 86 		& 141	\\
L4 	& 2.13 (0.19) 	& 50 		& 92	\\
L5 	& 2.16 (0.07) 	& 48 		& 90	\\
L6 	& 2.43 (0.23) 	& 30 		& 64	\\
L7 	& 2.59 (0.19) 	& 25 		& 56	\\
L8 	& 2.81 (0.21) 	& 18 		& 45 
\enddata
\tablenotetext{a}{Median $i-z$ colors and standard deviations calculated from \citet{west:2011:97} and \citet{schmidt:2015:158}.}
\tablenotetext{b}{Corresponding to $i = 21.3$ using the $i-z$ photometric parallax relationship from S. J. Schmidt et al. (2016, in preparation).}
\tablenotetext{c}{Corresponding to $z = 20.5$ using the $i-z$ photometric parallax relationship from S. J. Schmidt et al. (2016, in preparation).}
\end{deluxetable}

\subsubsection{2MASS and \emph{WISE} Matches}

	All of the SDSS objects were cross-matched to the nearest 2MASS and \emph{WISE} source within 8\arcsec. We matched to 2,364,776 objects in 2MASS, and 3,473,521 in \emph{WISE}. Of these matches, 1,741,279 were found in both 2MASS and \emph{WISE}. The SEDs of VLM objects with spectral types of M and L peak in the NIR. SDSS and \emph{WISE} are deeper surveys than 2MASS by approximately a factor of 4 in limiting flux \citep[see Table 1;][]{theissen:2016:41}, therefore M and L dwarfs detected by both SDSS and \emph{WISE} should be detected by 2MASS. For this purpose, we kept only objects that had matches in both 2MASS and \emph{WISE}. Additionally, we required 2MASS sources to be detected in two of the three bands to reduce spurious detections (1,654,053 objects). We further removed objects close to the Galactic plane ($|b| < 20^\circ$) which are prone to source confusion, especially for the large \emph{WISE} point-spread-functions (PSFs; 6.1\arcsec\ FHWM at 3.4 \um). This reduced the sample to 401,086 objects

\subsubsection{Neighbors}

	The largest source of contamination comes from neighboring objects with small angular separations from our primary object, primarily due to the large size of the \emph{WISE} beam (6.1\arcsec\ FHWM at 3.4 \um). Neighboring objects with small angular separations ($\lesssim$5\arcsec) may be resolved in SDSS and 2MASS, but are blended in \emph{WISE}, causing a shift in the measured astrometric position. \citet{theissen:2016:41} made a cut dependent on both neighbor distance (angular separation), and $r$-band contrast ratio between the primary object and neighboring object. Since our objects are intrinsically fainter than the objects in the original MoVeRS sample, we chose to make a full cut on neighbor distance, independent of $r$-band contrast ratio. We applied the same cut as \citet{theissen:2016:41}, removing any object that had a neighboring SDSS primary object within 8\arcsec. This left us with 98,165 objects found in SDSS, 2MASS, and \emph{WISE} with no detected neighbors within 8\arcsec.

\subsubsection{Proper Motions}\label{PMs}

\begin{figure*}
\centering
\includegraphics[width=\linewidth]{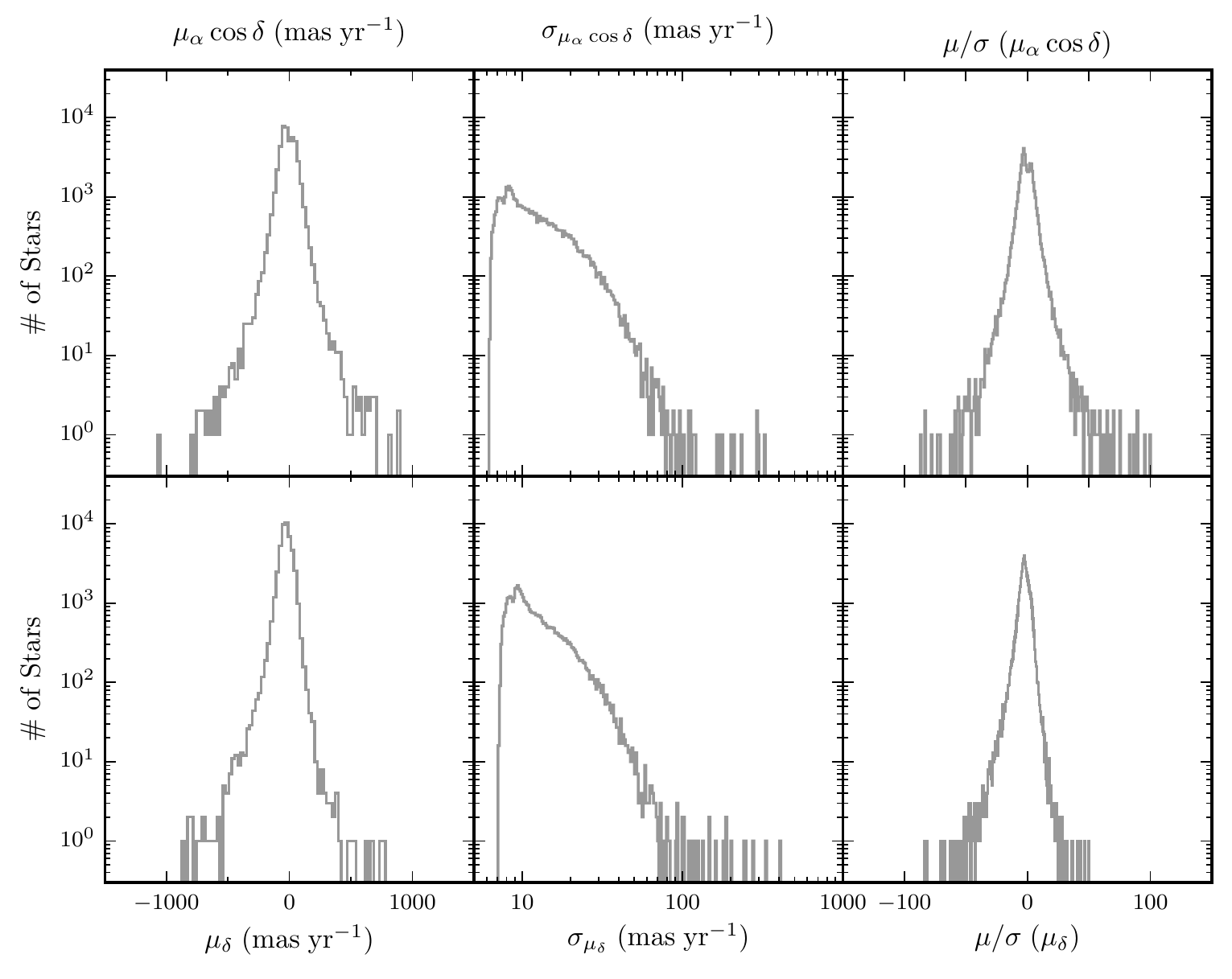}
\caption{Proper motion and proper motion error distributions for the LaTE-MoVeRS sample. The vast majority of our objects have $|\mu_\mathrm{tot}| < 1000$ mas yr$^{-1}$. Proper motion uncertainties for the LaTE-MoVeRS sample peak at values $< 10$ mas yr$^{-1}$. The inclusion of three epochs allows us to reduce the uncertainty from measurements made based solely on 2MASS and \emph{WISE} astrometry.
\label{fig:PMs}}
\end{figure*}

	Using all three epochs of astrometric data (SDSS, 2MASS, and \emph{WISE}), proper motions were computed with a weighted linear fit to the positions versus time using Orthogonal Distance Regression \citep{boggs:1990:183}. Further details on our methods for computing proper motions can be found in \citet{theissen:2016:41}. We kept objects with significant proper motions ($\mu_\mathrm{tot} \geqslant 2\sigma_{\mu_\mathrm{tot}}$), yielding \latemovers\ sources. The distributions of proper motions and proper motion errors for the LaTE-MoVeRS sample are shown in Figure~\ref{fig:PMs}. The proper motion and uncertainty distributions are very similar to the MoVeRS distributions computed with all three epochs (SDSS, 2MASS, and \emph{WISE}), as is expected. Our catalog has median uncertainties of 10.3 and 10.8 mas yr$^{-1}$ for $\sigma_{\mu_\alpha \cos \delta}$ and $\sigma_{\mu_\delta}$, respectively, providing a similar precision to that of the original MoVeRS catalog \citep[$\sigma_{\mu_\mathrm{tot}} \approx 10$ mas yr$^{-1}$;][]{theissen:2016:41}. Independent verifications to test the fidelity of the LaTE-MoVeRS proper motions are performed in the following section.

\subsection{Proper Motion Fidelities}\label{PMsFid}

	We tested the validity of our proper motion measurements by comparing them, where available, to other proper motion catalogs with reliable measurements. This is complicated by the fact that our catalog is reaching into a color and magnitude range that surpasses the limits of most previous catalogs. Even so, we can compare to the reddest proper motion catalogs available, which include the previous MoVeRS catalog, the Positions and Proper Motions Catalog \citep[PPMXL;][]{roser:2010:2440}, the SUPERBLINK survey \citep{lepine:2005:1483,lepine:2011:138}, the 1st United States Naval Observatory (USNO) Robotic Astrometric Telescope catalog \citep[URAT1;][]{zacharias:2015:101}, and the Bayesian Analysis for Nearby Young AssociatioNs \citep[BANYAN;][]{malo:2013:88} All-Sky Survey (BASS) catalog \citep{gagne:2015:73}.
	
	Our first comparison was to the original MoVeRS catalog \citep{theissen:2016:41}. We found 29,464 cross-matches between the two catalogs, the bulk of our LaTE-MoVeRS sample. Although there is significant overlap between the two samples, the methods used in reconstructing SDSS $i$-band astrometry yields more precise and reliable proper motion measurements for these redder objects. Comparing the agreement between both catalogs at the 1- and 2-$\sigma$ level for both proper motion components of the LaTE-MoVeRS sample shows very good agreement (99.82\% at the 1$\sigma$ level), as is listed in Table~\ref{tbl:agreement}.
	
\begin{figure}
\centering
 \includegraphics[width=\linewidth]{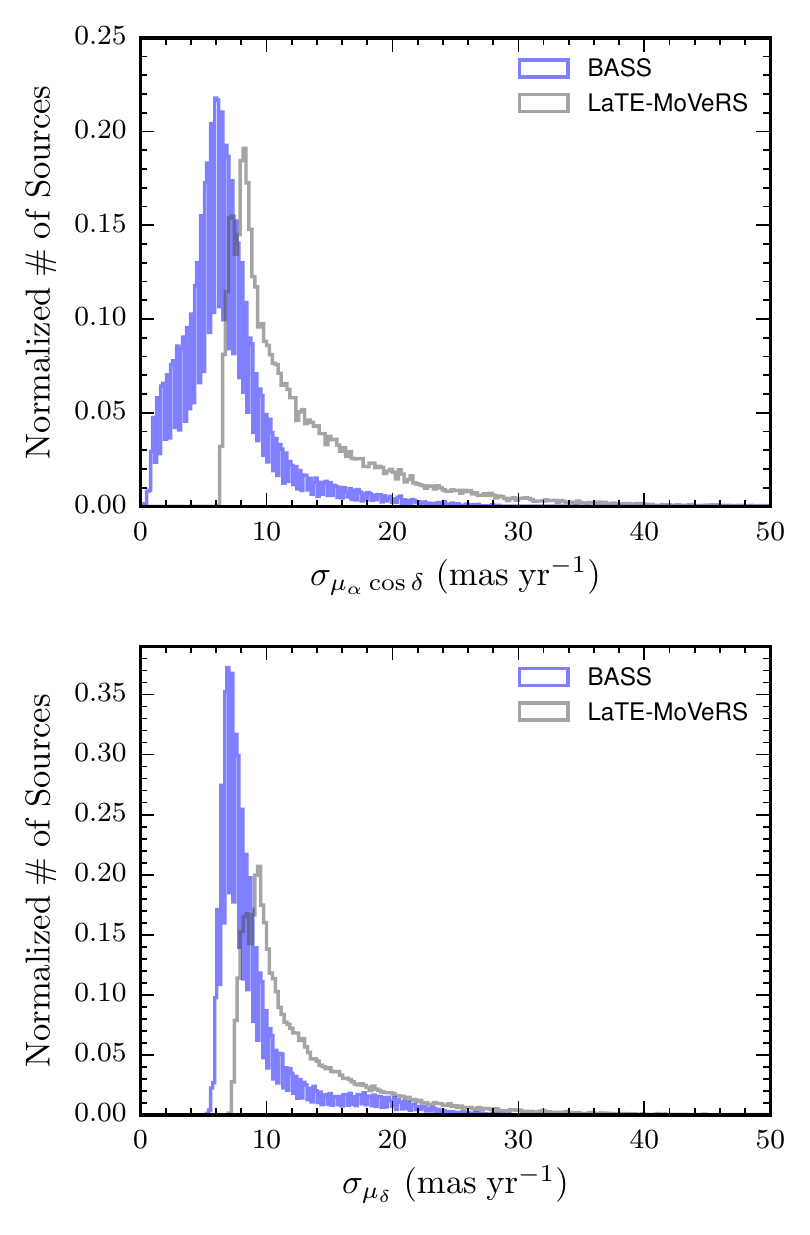}
\caption{Comparison of proper motion component uncertainties for BASS and the LaTE-MoVeRS sample. The systematically smaller $\sigma_{\mu_\alpha \cos \delta}$ values for BASS are due to the inclusion of a 2nd multiplicative factor of $\cos \delta$ \protect\citep[see Equation 1;][]{gagne:2015:73}. Another contributing factor to the larger uncertainties in LaTE-MoVeRS is the addition of a systematic uncertainty measured using QSO positions among 2MASS, SDSS, and \emph{WISE} \citep[for further details see][]{theissen:2016:41}. No such systematic offset was measured for the BASS sample.
\label{fig:BASScompare}}
\end{figure}

	The BASS catalog compiled by \citet{gagne:2015:73} is the ideal benchmark test for our proper motions since it was built using similar datasets (2MASS and \emph{WISE}). Although both LaTE-MoVeRS and BASS were searching for very-low-mass objects, the BASS catalog was designed specifically to look for young, nearby VLM stars and brown dwarfs associated with nearby young moving groups (NYMGs), whereas LaTE-MoVeRS makes no age discrimination. BASS required a multitude of extra selection criteria, specifically numerous color cuts (e.g., $J-K_s > 0.775$; for further details see \citealt{gagne:2015:73}). The BASS catalog contains 98,970 objects distributed over the entire sky. The BASS catalog claims a precision of 5--25 mas yr$^{-1}$, however, the respective astrometric errors for 2MASS and \emph{WISE} (at the bright end) are 60 mas and 90 mas, respectively. Combined with the $\sim$13 year maximum baseline between the two surveys, this translates into a lower uncertainty limit of $\sim$9 mas yr$^{-1}$. This uncertainty does not include the systematic astrometric uncertainty between the two surveys, which may be on the order of 50 mas or more \citep{theissen:2016:41}. A comparison of the proper motion uncertainties for BASS and LaTE-MoVeRS is shown in Figure~\ref{fig:BASScompare}. The systematically lower values of $\sigma_{\mu_\alpha \cos \delta}$ in BASS are due to a 2nd multiplicative factor of $\cos \delta$ in Equation (1) of \citet{gagne:2015:73}, which is already included in the 2MASS astrometric uncertainties (Vandana Desai, personal communication). The LaTE-MoVeRS proper motion uncertainties should be more robust than those reported in BASS.
	
	Uncertainties aside, the measured values should be in very close agreement to our derived proper motions for stars with $\mu > 30$ mas yr$^{-1}$ (the BASS limit). The addition of the SDSS epoch for the LaTE-MoVeRS proper motions results in higher-precision proper motions. This effectively makes LaTE-MoVeRS not only more precise than BASS, but also deeper for our earliest type sources, as more distant objects will exhibit smaller tangential motions. We matched 6,990 of our objects to the BASS catalog and found very good agreement between proper motion measurements within their respective errors (99.93\% at the 1$\sigma$ level; see Table~\ref{tbl:agreement}). The large amount of LaTE-MoVeRS sources missing from BASS (39,520 sources) is primarily due to the more stringent selection criteria implemented for BASS. Using only the color selection criteria from BASS, specifically $0.506 < J-H < 2$, $0.269 < H-K_s < 1.6$, and $0.168 < W1-W3 < 2.5$, reduces the LaTE-MoVeRS sample to 18,819 sources. However, the use of significant proper motions to select VLM stars and brown dwarfs does not motivate the need for such stringent criteria.
	
	URAT1 has a $\sim$3 year baseline, a wide field-of-view (28 deg.$^2$), and uses a red bandpass (6800--7620\AA), making it more suitable for faint, low-mass objects than surveys with longer baselines, smaller fields-of-view, and/or bluer bandpasses (e.g., the USNO-B1.0 catalog, \citealt{monet:2003:984} and 4th USNO CCD Astrograph Catalog, UCAC4 \citealt{zacharias:2013:44}). The precision of URAT1 is between 5--7 mas yr$^{-1}$ \citep{zacharias:2015:101}. We cross-matched 15,390 of our objects to the URAT1 catalog, again finding extremely good agreement between the two measurements to within errors (see Table~\ref{tbl:agreement}).
	
	Using the SUPERBLINK software \citep{lepine:2002:1190,lepine:2003:921}, \citet[][hereafter LSPM]{lepine:2005:1483} completed a reanalysis of the Digitized Sky Surveys (DSS) to search for high-proper motion stars. Although the shallow magnitude limit of DSS precludes a large number of matches to the LaTE-MoVeRS catalog, the high-fidelity of the LSPM catalog (all candidates verified by-eye) makes it a useful. Although formal errors are not stated for objects in LSPM, the dispersion of LSPM proper motions compared to UCAC2 \citep{zacharias:2004:3043} is $\sim$7 mas yr$^{-1}$. Comparing the 808 stars matched between LaTE-MoVeRS and LSPM, we find greater than 98\% agreement at the 2$\sigma$ level (see Table~\ref{tbl:agreement}). 
	
	Our final comparison was made between the PPMXL catalog \citep{roser:2010:2440}, built using the USNO-B1.0 survey and 2MASS. To select the highest-quality sources from PPMXL, we required an object to be detected in $>$ 4 epochs, and have no quality flags raised. This produced 6,855 cross-matched sources, with relatively good agreement (88.84\% at the 2$\sigma$ level; see Table~\ref{tbl:agreement}), although not as high as our comparisons with other proper motion catalogs. There are a large number of high proper motion sources in PPMXL that do not have similarly high-proper motions in LaTE-MoVeRS. \citet{roser:2010:2440} note the large number of high proper motion sources ($>$150 mas yr$^{-1}$) in the northern hemisphere of PPMXL, and caution that these are likely spurious measurements within PPMXL. The high agreement of LaTE-MoVeRS with other proper motion samples appears to confirm this. Our comparisons indicate that the proper motions of LaTE-MoVeRS are reliable, and our uncertainty estimates are robust.
	
\begin{deluxetable}{c c c r}
\tabletypesize{\footnotesize}
\tablecolumns{4}
\tablecaption{Proper Motion Agreement\label{tbl:agreement}}
\tablehead{
\colhead{Catalog} & \colhead{Number of} & \multicolumn{2}{c}{Agreement} \\
\colhead{} & \colhead{Matches} & \colhead{1$\sigma$} & \colhead{2$\sigma$}
}
\startdata
MoVeRS 					& 29464	& 99.82\%	& 99.98\%\\
BASS					& 6990	& 99.93\%	& 100.00\%\\
URAT1					& 15390	& 88.68\%	& 98.10\%\\
LSPM 					& 808	& 80.44\%	& 98.76\%\\
PPMXL			 		& 6855	& 64.64\%	& 88.84\%
\enddata
\end{deluxetable}

\section{Quantifying the Gaia Shortfall}\label{Gaia}

	Currently, the largest astrometric mission to obtain positions and proper motions is underway by \emph{Gaia}. \emph{Gaia} is expected to obtain parallaxes and proper motions, among other measurements, for over a billion stars with extremely small parallax uncertainties \citep[$\lesssim$16 $\mu$as for low-mass stars with $V \lesssim 14$;][]{perryman:2001:339}. Recently, \emph{Gaia} Data Release 1 \citep[DR1;][]{gaia-collaboration:2016:a2} was made public, including five parameter astrometry solutions (positions, parallaxes, and proper motions) for stars already included in \emph{Hipparcos} \citep{ESA:1997:1} and Tycho-2. The precision of DR1 for the five-parameter astrometry solution for sources with $G \approx 12$ is estimated to be 0.75 mas for positions, 0.64 mas for parallaxes, and 3.19 mas yr$^{-1}$ for proper motions. However, due to the faintness of our sources, none are included in the subset of DR1 with five-parameter astrometry. DR1 also contains the positions and $G$-band (\emph{Gaia} passband) magnitudes for 1,142,679,769 sources. Although the \emph{Gaia} mission will culminate in the largest astrometric catalog to date with positions and proper motions, this catalog is estimated to be incomplete for VLM stars and brown dwarfs in comparison to what is achievable with SDSS+2MASS+\emph{WISE} \citep{theissen:2016:41}. The spatial resolution of \emph{Gaia} \citep[$\sim$50 mas pixel$^{-1}$;][]{perryman:2001:339} is an order of magnitude better than what SDSS achieves. Therefore, the primary factor in limiting \emph{Gaia} for studies of VLM stars and brown dwarfs (in comparison to SDSS+2MASS+\emph{WISE}) is \emph{Gaia}'s relatively blue bandpass \citep{perryman:2001:339}, and the magnitude limit of the astrometric instrument \citep[$r \lesssim 20$;][]{ivezic:2012:251}, making it shallower than the combined SDSS+\emph{WISE} dataset \citep[$r \lesssim 22.2$;][]{theissen:2016:41}.
	
\begin{figure*}
\centering
\includegraphics{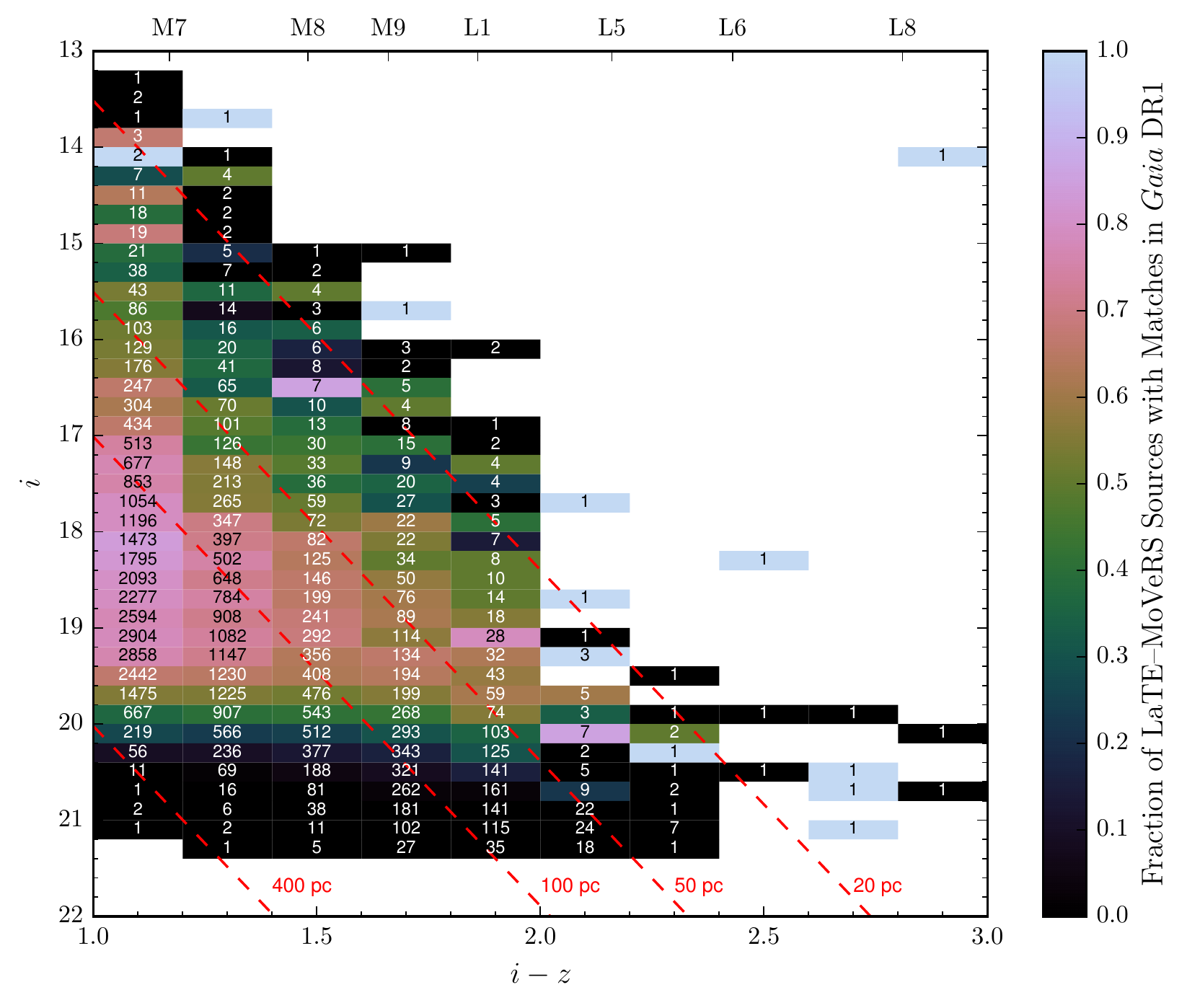}
\caption{Map of the number of LaTE-MoVeRS sources with counterparts in \emph{Gaia} DR1 as a function of $i-z$ color and $i$-band magnitude. The color of the bin corresponds to the fraction of stars with matches in \emph{Gaia} DR1, and the number within the bin indicates the total number of LaTE-MoVeRS sources within that bin (text colored black indicates a fraction $\geqslant 0.7$). We also show expected magnitudes and colors corresponding to different distances (red dashed lines) using the S. J. Schmidt et al. (2016, in preparation) $i-z$ photometric parallax relationship (Section~\ref{distancetemp}). The fraction of \emph{Gaia} DR1 matches drops below $\sim$30\% for all colors (spectral types) with $i \geqslant 20$. Approximate spectral types are shown from Table~\ref{tbl:colors}. The two outliers are discussed in the text.
\label{fig:Gaia1}}
\end{figure*}

	With DR1 now released, we compared the \emph{Gaia} sources to LaTE-MoVeRS to empirically estimate the incompleteness of VLM stars and brown dwarfs within the \emph{Gaia} catalog. We matched the \latemovers\ sources within the LaTE-MoVeRS catalog to the nearest source within 8\arcsec\ in \emph{Gaia} DR1, producing 29,535 matches. The $i-z$ and $i$-band magnitude map of the fraction of LaTE-MoVeRS sources matched to \emph{Gaia} DR1 are shown in Figure~\ref{fig:Gaia1}. The fraction of matches to \emph{Gaia} DR1 falls below $\sim$30\% for all sources with $i \geqslant 20$. For sources with $i \leqslant 20$, the weighted average fraction of matches is $\sim$68\%, indicating that even for brighter sources \emph{Gaia} will miss a large fraction of VLM stars and brown dwarfs within the combined SDSS+2MASS+\emph{WISE} dataset. There is a slight dependence on the fraction of matches with $i-z$ color (spectral type), which is expected as redder objects will have little flux in the relatively blue \emph{Gaia} bandpass. 

	There are two outliers detached from the bulk of the population in Figure~\ref{fig:Gaia1}. The first outlier at $i-z \approx 2.5$, $i \approx 18.3$ (SDSS J141624.08+134826.7) is a known L5 dwarf \citep{bowler:2010:45, schmidt:2010:1045} at $\sim$10 pc \citep{kirkpatrick:2012:156}, and the primary component of a resolved binary system with a T7.5 secondary \citep{burningham:2010:1952, burgasser:2010:2448, scholz:2010:l8}. The second outlier at $i-z \approx 2.9$, $i \approx 14.1$ (2MASS J17081033$-$0220225) has been photometrically identified as an asymptotic giant branch (AGB) star and a weak SiO maser source \citep{ita:2001:112, deguchi:2010:525}. This star was also targeted for a study of H$_2$O masers \citep{yung:2013:20}, but was considered a non-detection. Additionally, this object has been identified as a long-period variable with a period of $\sim$385 days \citep{drake:2014:9}, consistent with the AGB phase. This star has a small detected proper motion in multiple catalogs \citep{monet:2003:984, roser:2010:2440, cutri:2014:, zacharias:2015:101}, a photometric distance assuming a VLM object (see Section~\ref{distancetemp}) would place this star at a distance of $\sim$1 pc. Given the observational evidence, this object is most likely a giant rather than a dwarf. The detected proper motion could be due to saturation within the \emph{WISE} passbands, which would result in a shifted photocenter. Source variability could also shift the measured photocenter \citep[e.g., variability-induced movers;][]{wielen:1996:679}.

	Figure~\ref{fig:Gaia1} indicates that \emph{Gaia} will only be able to detect objects with spectral types of $\sim$L3 out to distances $< 50$ pc, and requires later spectral types to be much closer. This limits the use of \emph{Gaia} for studies of VLM stars and brown dwarfs, and necessitates the use of datasets such as SDSS, 2MASS, and \emph{WISE} for further identification of the lowest-mass objects out to larger volumes.

\section{The LaTE-MoVeRS Sample}\label{latemovers}

\begin{figure}
\centering
\includegraphics{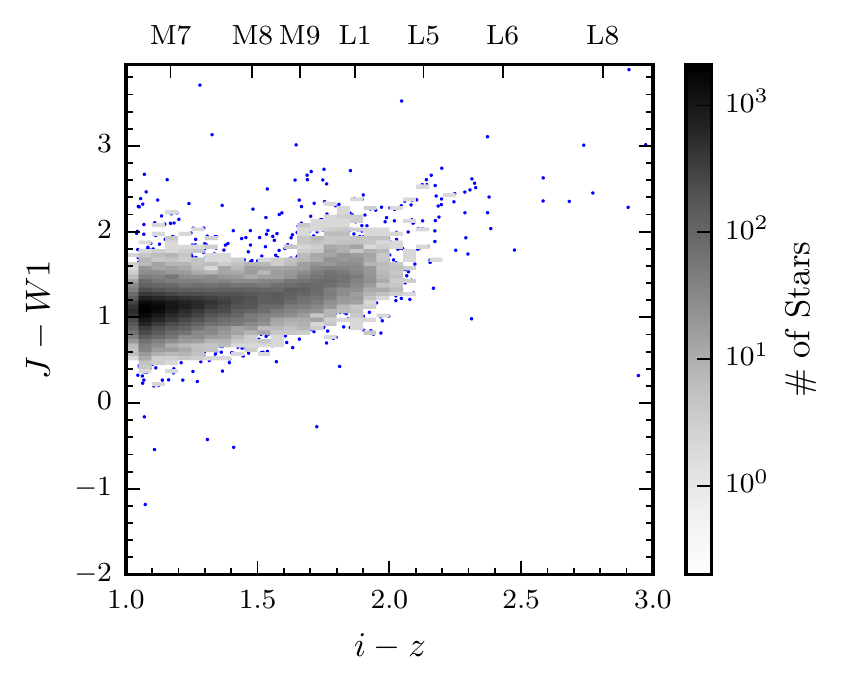}
\caption{2D histogram of the color distributions for the LaTE-MoVeRS sample. Each bin is (0.05 mag)$^2$. Single stars are shown as points. Also shown are approximate spectral types from \protect\citet{schmidt:2010:1808,schmidt:2015:158}. Although the LaTE-MoVeRS sample is primarily composed of late-type M dwarfs, the sample extends down to late L-type objects.
\label{fig:colordist}}
\end{figure}

	The LaTE-MoVeRS sample was selected to extend the MoVeRS sample to later spectral types. The selection criteria preferentially select for dwarfs with spectral types later than M5. Figure~\ref{fig:colordist} shows the color distribution of our sample in SDSS, 2MASS, and \emph{WISE} colors. Approximate spectral types from \citet{schmidt:2010:1808,schmidt:2015:158} are shown.

\begin{figure}
\centering
 \includegraphics[width=\linewidth]{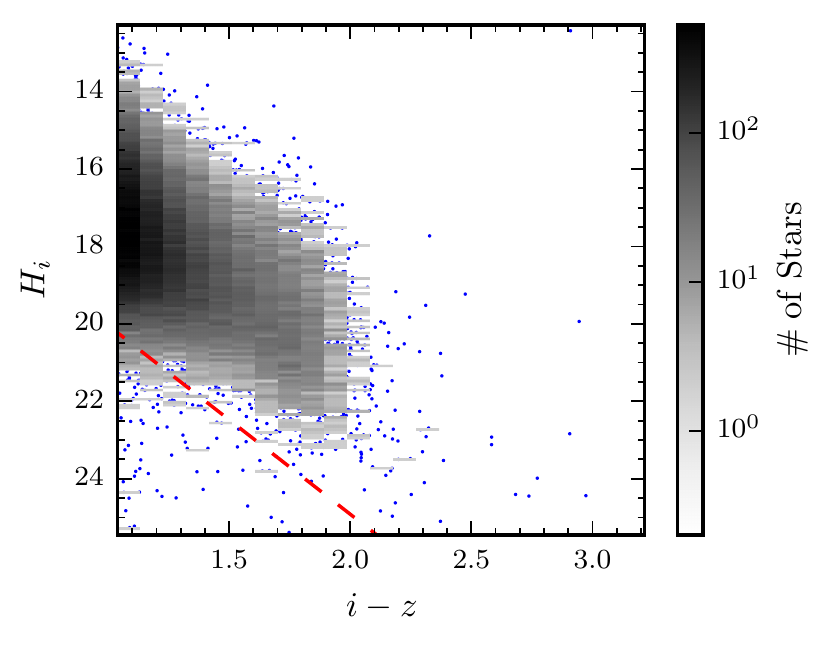}
\caption{Reduced proper motion diagram for the LaTE-MoVeRS sample, each bin is (0.1 mags)$^2$. Single blue points mark individual objects. The dashed line represent a tangential velocity of 180 km s$^{-1}$, which separates disk stars from halo stars \protect\citep{sesar:2008:1244, dhital:2010:2566}. The kinematics of the LaTE-MoVeRS sample are consistent with a population dominated by disk objects.
\label{fig:rpm}}
\end{figure}

	To illustrate the mix of thin disk (young) and thick disk (old) dwarfs within the LaTE-MoVeRS sample, we plot the $i$-band reduced proper motions, defined as $H_i = i + 5 + 5 \log\mu = M_i - 3.25 + 5 \log v_T$ \citep{luyten:1922:135}, as a function of $i-z$ color in Figure~\ref{fig:rpm}. We also show the expected trend for objects with tangential velocities of 180 km s$^{-1}$ (red dashed line), the demarcation point between disk and halo objects \citep{sesar:2008:1244, dhital:2010:2566}. Although halo objects can be found with tangential velocities below this limit, disk objects are rarely found with tangential velocities above this limit \citep{theissen:2016:41}. The majority of the LaTE-MoVeRS sample appears to be made up of disk objects.

\subsection{Photometric Distances and Stellar Temperatures}\label{distancetemp}

	M and L dwarfs near our solar system ($<50$ parsecs) make ideal candidates for numerous astrophysical investigations. These nearby objects are used to study NYMGs \citep[e.g.,][]{shkolnik:2012:56}, Earth-sized planets \citep[e.g.,][]{berta:2012:145}, unresolved binaries \citep[e.g.,][]{faherty:2010:176,best:2013:84}, wide binaries \citep[e.g.,][]{burgasser:2012:110, bardalez-gagliuffi:2014:143}, and the local mass function \citep[e.g.,][]{winters:2015:5}. Accurate distances from trigonometric parallaxes are very costly in terms of time and precision, requiring multiple epochs to accurately determine parallaxes. From Section~\ref{Gaia}, it is clear \emph{Gaia} will not obtain astrometric measurements (parallaxes and proper motions) for objects with spectral types later than $\sim$L3 out to distances of $\sim$50 pc (down to less than 20 pc for objects with spectral types later than L5). \emph{WISE} is able to detect VLM stars and brown dwarfs out to distances surpassing the distance limits of \emph{Gaia}. However, measuring accurate parallaxes using \emph{WISE} multi-epoch photometry is difficult due to the cadence of observations and large PSFs, making astrometric measurements a blend of proper motion and parallax or \emph{apparent motion} \citep{kirkpatrick:2014:122}. Parallax measurements with \emph{Spitzer} can be achieved, however, observing campaigns can be costly, and only a small number of parallaxes have been measured for T and Y dwarfs \citep{dupuy:2013:1492}.
	
	With no follow-up space-based astrometric mission planned to measure astrometry for the VLM stars and brown dwarfs that reside within the SDSS+2MASS+\emph{WISE} dataset, ground-based observations can extend the census of VLM stars and brown dwarfs past the limitations of \emph{Gaia}. Efforts are underway to obtain trigonometric parallax measurements of VLM stars and brown dwarfs within 33 pc in the SDSS footprint (J. Skinner et al. 2017 in preparation), which can be used to further refine photometric parallax relations in SDSS passbands for future telescopes such as the Large Synoptic Survey Telescope \citep[LSST;][]{ivezic:2008:} and the southern hemisphere SDSS. Currently, few VLM stars and brown dwarfs have both measured parallaxes and photometry in SDSS passbands.

	Using 65 M and L Dwarfs with parallax measurements, SDSS photometry, and previously classified spectral types, S. J. Schmidt et al. (2016, in preparation) derived photometric parallax relations for $M_i$ as a function of the $i-z$, $i-J$ and $i-K$ colors with typical uncertainties on the calculated distances better than 20\% for all three relations. S. J. Schmidt et al. (2016, in preparation) compared distances using the derived photometric parallax relations to distances calculated with photometric parallax relations from \citet{bochanski:2010:2679}, \citet{schmidt:2010:1808} and \citet{dupuy:2012:19}. S. J. Schmidt et al. (2016, in preparation) found that 80--95\% of the distances agreed within the 1-$\sigma$ uncertainty for each measurement in all three comparisons.
	
\begin{figure}
\centering
\includegraphics[width=\linewidth]{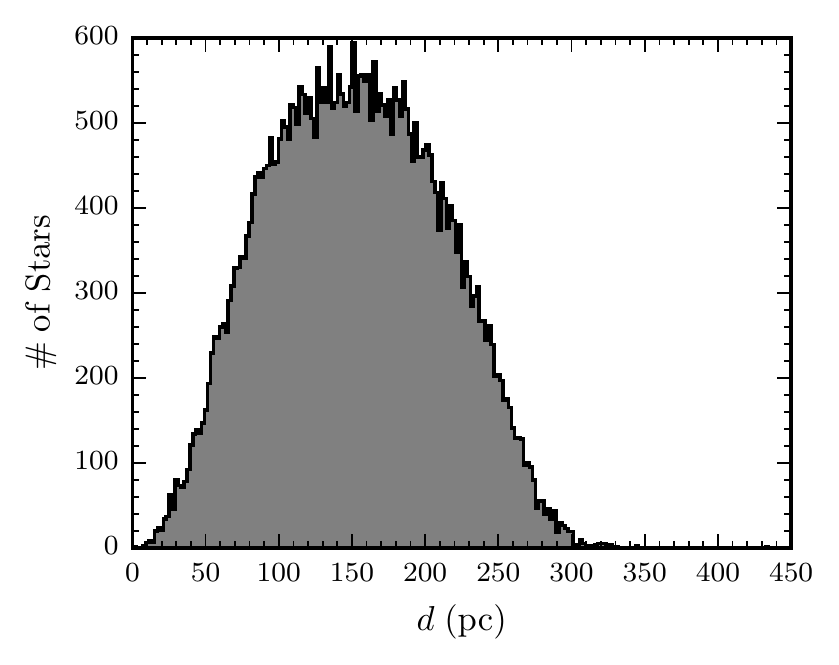}
\caption{Distance distribution for the LaTE-MoVeRS sample using the S. J. Schmidt et al. (2016, in preparation) photometric parallax relations.
\label{fig:dists}}
\end{figure}

	We computed distances for the LaTE-MoVeRS sample using all S. J. Schmidt et al. (2016, in preparation) relationships where available. We then computed an uncertainty weighted mean distance using all available distances. The distribution of distances for the LaTE-MoVeRS sample is shown in Figure~\ref{fig:dists}. Of the 171 objects with estimated distances $\leqslant$ 25 pc, 13 have not been previously identified as nearby VLM stars or brown dwarfs according to VizieR\footnote{\url{http://vizier.u-strasbg.fr/}}, SIMBAD\footnote{\url{http://simbad.u-strasbg.fr/simbad/}}, and the Astrophysics Data System\footnote{\url{http://adswww.harvard.edu/}} (ADS). We list the details for these 13 nearby systems in Table~\ref{tbl:nearby}. Three of the 13 objects have spectra available through SDSS (one with a BOSS optical spectrum and two with APOGEE near-infrared spectra). We estimated a spectral type of dM8 for the object with the BOSS spectrum (SDSS DR8$+$ objID 1237655124466401512; 2MASS J11230124+0400411) using the \emph{PyHammer} code \citep{kesseli:2016:}.

\begin{deluxetable*}{lrrlrrcc}
\tabletypesize{\footnotesize}
\tablecolumns{8}
\tablecaption{Newly Identified Nearby VLM Objects\label{tbl:nearby}}
\tablehead{
\colhead{SDSS ObjID} & \colhead{R.A.} & \colhead{Dec.} & \colhead{$T_\mathrm{eff}$} & \colhead{$\mu_\alpha$} & \colhead{$\mu_\delta$} & \colhead{$d$} & \colhead{Spectrum} \\
\colhead{DR8$+$} & \colhead{(deg.)} & \colhead{(deg.)} & \colhead{$(K)$} & \colhead{(mas yr$^{-1}$)} & \colhead{(mas yr$^{-1}$)} & \colhead{(pc)} & \colhead{Available?\tablenotemark{a}}
}
\startdata
1237652901822529699 	& 1.651612 	& $-$9.100849 	& $2986^{+28}_{-89}$	& $-170 \pm 25$	& $8 \pm 8$	& $23 \pm 2$	& N\\
1237673702350782616 	& 33.071576	& $-$10.608396 & $2599^{+19}_{-18}$	& $-54 \pm 11$		& $26 \pm 8$	& $24 \pm 2$	& N\\
1237667228233236692\tablenotemark{b} 	& 56.781736	& 4.298350 	& $2498^{+20}_{-22}$	& $120 \pm 13$	& $-37 \pm 34$	& $21 \pm 2$	& N\\
1237667206217072893\tablenotemark{b}	& 123.776238	& 65.366125	& $3002^{+20}_{-19}$	& $-42 \pm 14$		& $-139 \pm 16$ & $25 \pm 2$	& N\\
1237667486456348855\tablenotemark{b} 	& 138.671756	& 18.162400 	& $2901^{+23}_{-21}$	& $-153 \pm 15$	& $-9 \pm 12$	& $23 \pm 2$	& N\\
1237655124466401512 	& 170.755215 	& 4.011425 	& $2596^{+20}_{-19}$	& $11 \pm 8$		& $-28 \pm 9$	& $24 \pm 3$	& 1\\
1237661433234784335\tablenotemark{b} 	& 185.459060	& 46.545672 	& $2795^{+21}_{-76}$	& $90 \pm 27$		& $7 \pm 14$	& $15 \pm 1$	& N\\
1237661966359789628 	& 195.320759	& 41.211235 	& $2906^{+90}_{-29}$	& $64 \pm 7$		& $-3 \pm 14$	& $23 \pm 2$	& 2\\
1237665532243542117\tablenotemark{b} 	& 200.974433	& 26.855418 	& $2702^{+31}_{-24}$	& $-114 \pm 11$	& $-18 \pm 10$	& $25 \pm 2$	& N\\
1237662193995939989\tablenotemark{b} 	& 203.998759	& 40.362813 	& $2977^{+32}_{-92}$	& $38 \pm 10$		& $26 \pm 8$	& $20 \pm 2$	& 2\\
1237671992963629135\tablenotemark{c}	& 217.766191	& $-$22.118923 & $2489^{+26}_{-112}$	& $23 \pm 23$		& $-45 \pm 10$	& $16 \pm 4$	& N\\
1237651250992514994\tablenotemark{b} 	& 265.963094 	& 18.549266 	& $2297^{+19}_{-20}$	& $-183 \pm 8$		& $-98 \pm 9$	& $24 \pm 3$	& N\\
1237656530534924602\tablenotemark{b} 	& 267.330274	& 47.934751 	& $2798^{+18}_{-18}$	& $-49 \pm 7$		& $-29 \pm 8$	& $17 \pm 2$	& N
\enddata
\tablenotetext{a}{$1 =$ SDSS optical spectrum available. $2 = $ SDSS APOGEE spectrum available.}
\tablenotetext{b}{Identified as a VLM star or brown dwarf in \citet{gagne:2015:73} but no distance was determined.}
\tablenotetext{c}{Identified photometrically as a potential quasar in \citet{richards:2015:39}.}
\end{deluxetable*}

\begin{figure}
\centering
 \includegraphics[width=\linewidth]{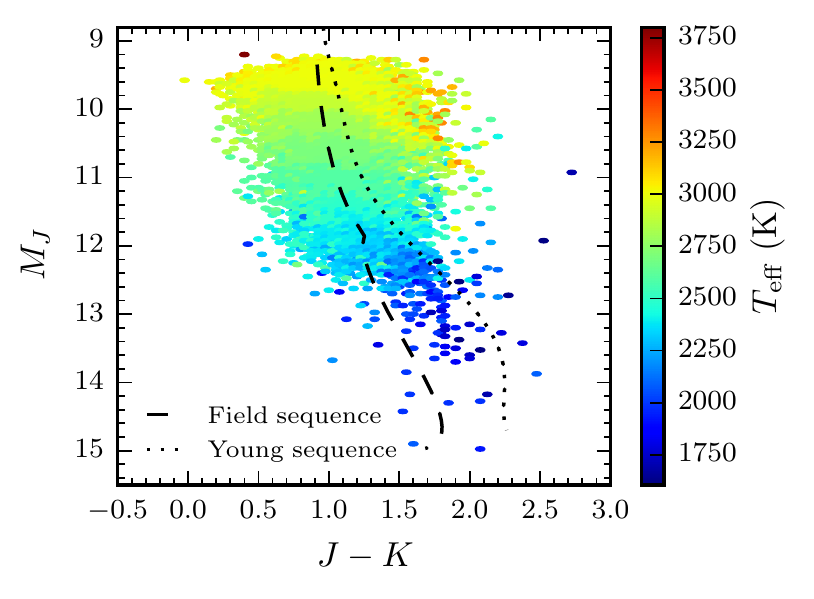}
\caption{Color--absolute magnitude diagram for the LaTE-MoVeRS sample using distances computed from the S. J. Schmidt et al. (2016, in preparation) relationships and $T_\mathrm{eff}$ estimates from our MCMC method. Each bin is (0.05)$^2$ mag$^2$, and displays the median $T_\mathrm{eff}$ within the bin. Also shown are the semi-empirical calibrations from \protect\citet{gagne:2015:33} for field objects and young objects (ages $\lesssim$ 100 Myr). The LaTE-MoVeRS sample shows the expected trend with temperature and luminosity. The LaTE-MoVeRS sample also appears to have a mix of both field objects and young objects.
\label{fig:cmd}}
\end{figure}

	The extensive wavelength coverage from SDSS, 2MASS, and \emph{WISE} allows us to estimate stellar effective temperatures based on the SEDs of our objects. We used the BT-Settl models \citep{allard:2012:2765, allard:2012:3} and sampled for a best-fit stellar photosphere with a Markov Chain Monte Carlo (MCMC) method using the \emph{emcee} Python package \citep{foreman-mackey:2013:306}. Details of our methods for estimating stellar parameters using the BT-Settl models and the \emph{emcee} are described in \citet{theissen:2016:2}. Combining our distance and $T_\mathrm{eff}$ estimates we show the color--absolute magnitude diagram of the LaTE-MoVeRS sample in Figure~\ref{fig:cmd}. The LaTE-MoVeRS catalog is available through the online journal, VizieR, and SDSS CasJobs, and the column descriptions are listed in Table~\ref{tbl:latemovers}.
	
\LongTables
\begin{deluxetable}{cll}
\tabletypesize{\footnotesize}
\tablecolumns{3}
\tablecaption{LaTE-MoVeRS Schema\label{tbl:latemovers}}
\tablehead{
\colhead{Column} & \colhead{Column} & \colhead{Units} \\
\colhead{Number} & \colhead{Description} & 
}
\startdata
1	& SDSS Object ID & ... \\
2	& SDSS R.A. & deg. \\
3	& SDSS Decl. & deg. \\
4	& SDSS R.A. error ($\Delta \cos\delta$) & deg. \\
5	& SDSS Decl. error & deg. \\
6	& SDSS MJD & day \\
7	& SDSS $u$-band PSF mag. & mag \\
8	& SDSS $u$-band PSF mag. error & mag \\
9	& SDSS $u$-band extinction & mag \\
10	& SDSS $u$-band unreddened PSF mag. & mag \\
11	& SDSS $g$-band PSF mag. & mag \\
12	& SDSS $g$-band PSF mag. error & mag \\
13	& SDSS $g$-band extinction & mag \\
14	& SDSS $g$-band unreddened PSF mag. & mag \\
15	& SDSS $r$-band PSF mag. & mag \\
16	& SDSS $r$-band PSF mag. error & mag \\
17	& SDSS $r$-band extinction & mag \\
18	& SDSS $r$-band unreddened PSF mag. & mag \\
19	& SDSS $i$-band PSF mag. & mag \\
20	& SDSS $i$-band PSF mag. error & mag \\
21	& SDSS $i$-band extinction & mag \\
22	& SDSS $i$-band unreddened PSF mag. & mag \\
23	& SDSS $z$-band PSF mag. & mag \\
24	& SDSS $z$-band PSF mag. error & mag \\
25	& SDSS $z$-band extinction & mag \\
26	& SDSS $z$-band unreddened PSF mag. & mag \\
27	& 2MASS R.A. & deg. \\
28	& 2MASS Decl. & deg. \\
29	& 2MASS R.A. error ($\Delta \cos\delta$) & deg. \\
30	& 2MASS Decl. error & deg. \\
31	& 2MASS MJD & day \\
32	& 2MASS photometric quality flag & ... \\
33	& 2MASS read flag & ... \\
34	& 2MASS blend flag & ... \\
35	& 2MASS contamination \& confusion flag & ... \\
36	& 2MASS extended source flag & ... \\
37	& 2MASS $J$-band PSF mag. & mag \\
38	& 2MASS $J$-band PSF corr. mag. unc. & mag \\
39	& 2MASS $J$-band PSF total mag. unc. & mag \\
40	& 2MASS $J$-band SNR & ... \\
41	& 2MASS $J$-band extinction & mag \\
42	& 2MASS $J$-band unreddened PSF mag. & mag \\
43	& 2MASS $H$-band PSF mag. & mag \\
44	& 2MASS $H$-band PSF corr. mag. unc. & mag \\
45	& 2MASS $H$-band PSF total mag. unc. & mag \\
46	& 2MASS $H$-band SNR & ... \\
47	& 2MASS $H$-band extinction & mag \\
48	& 2MASS $H$-band unreddened PSF mag. & mag \\
49	& 2MASS $K_s$-band PSF mag. & mag \\
50	& 2MASS $K_s$-band PSF corr. mag. unc. & mag \\
51	& 2MASS $K_s$-band PSF total mag. unc. & mag \\
52	& 2MASS $K_s$-band SNR & ... \\
53	& 2MASS $K_s$-band extinction & mag \\
54	& 2MASS $K_s$-band unreddened PSF mag. & mag \\
55	& 2MASS $J$-band $\rchi^2_\nu$ goodness-of-fit & ... \\
56	& 2MASS $H$-band $\rchi^2_\nu$ goodness-of-fit & ... \\
57	& 2MASS $K_s$-band $\rchi^2_\nu$ goodness-of-fit & ... \\
58	& \emph{WISE} R.A. & deg. \\
59	& \emph{WISE} Decl. & deg. \\
60	& \emph{WISE} R.A. error ($\Delta \cos\delta$) & deg. \\
61	& \emph{WISE} Decl. error & deg. \\
62	& \emph{WISE} contamination \& confusion flag & ... \\
63	& \emph{WISE} extended source flag & ... \\
64	& \emph{WISE} variability flag & ... \\
65	& \emph{WISE} photometric quality flag & ... \\
66	& \emph{WISE} $W1$-band average MJD & day \\
67	& \emph{WISE} MJD uncertainty\tablenotemark{a} & day \\
68	& \emph{WISE} $W1$-band PSF mag. & mag \\
69	& \emph{WISE} $W1$-band PSF mag. unc. & mag \\
70	& \emph{WISE} $W1$-band SNR & ... \\
71	& \emph{WISE} $W1$-band extinction & mag \\
72	& \emph{WISE} $W1$-band unreddened PSF mag. & mag \\
73	& \emph{WISE} $W1$-band $\rchi^2_\nu$ goodness-of-fit & ... \\
74	& \emph{WISE} $W2$-band PSF mag. & mag \\
75	& \emph{WISE} $W2$-band PSF mag. unc. & mag \\
76	& \emph{WISE} $W2$-band SNR & ... \\
77	& \emph{WISE} $W2$-band extinction & mag \\
78	& \emph{WISE} $W2$-band unreddened PSF mag. & mag \\
79	& \emph{WISE} $W2$-band $\rchi^2_\nu$ goodness-of-fit & ... \\
80	& \emph{WISE} $W3$-band PSF mag. & mag \\
81	& \emph{WISE} $W3$-band PSF mag. unc. & mag \\
82	& \emph{WISE} $W3$-band SNR & ... \\
83	& \emph{WISE} $W3$-band extinction & mag \\
84	& \emph{WISE} $W3$-band unreddened PSF mag. & mag \\
85	& \emph{WISE} $W3$-band $\rchi^2_\nu$ goodness-of-fit & ... \\
86	& \emph{WISE} $W4$-band PSF mag. & mag \\
87	& \emph{WISE} $W4$-band PSF mag. unc. & mag \\
88	& \emph{WISE} $W4$-band SNR & ... \\
89	& \emph{WISE} $W4$-band extinction & mag \\
90	& \emph{WISE} $W4$-band unreddened PSF mag. & mag \\
91	& \emph{WISE} $W4$-band $\rchi^2_\nu$ goodness-of-fit & ... \\
92	& Proper motion in R.A. ($\mu_\alpha \cos \delta$) & mas yr$^{-1}$ \\
93	& Proper motion in Decl. ($\mu_\alpha \cos \delta$) & mas yr$^{-1}$ \\
94	& Intrinsic error in R.A. proper motion & mas yr$^{-1}$ \\
95	& Intrinsic error in Decl. proper motion & mas yr$^{-1}$ \\
96	& Fit error in R.A. proper motion & mas yr$^{-1}$ \\
97	& Fit error in Decl. proper motion & mas yr$^{-1}$ \\
98	& Total error in R.A. proper motion & mas yr$^{-1}$ \\
99	& Total error in Decl. proper motion & mas yr$^{-1}$ \\
100	& Proper motions time baseline & year \\
101	& Photometric distance & pc \\
102	& Photometric distance unc. & pc \\
103	& $T_\mathrm{eff}$ estimate & K \\
104	& Upper $T_\mathrm{eff}$ limit & K \\
105	& Lower $T_\mathrm{eff}$ limit & K \\
106	& Log $g$ estimate & dex \\
107	& Upper Log $g$ limit & dex \\
108	& Lower Log $g$ limit & dex 
\enddata
\tablenotetext{a}{Defined as $.5 \times $(W1MJDMAX$-$W1MJDMIN).}
\end{deluxetable}

	We acknowledge that many of our stars are outside the local bubble (distances $\gtrsim 60$ pc), and may suffer from extinction effects due to dust. However, the requirement that sources be located at relatively high Galactic latitudes ($|b| > 20^\circ$) minimizes reddening due to interstellar dust. SDSS provides extinction estimates for each bandpass ($A_\lambda$) using the \emph{Infrared Astronomical Satellite} \citep[\emph{IRAS};][]{neugebauer:1984:l1} 100 \um\ dust maps created by \citet[][hereafter SFD]{schlegel:1998:525}. These values estimate the total extinction along a line-of-sight out of the Galaxy, and may overestimate the actual extinction to our sources. The vast majority of our LaTE-MoVeRS sources (89\%) have $A_i < 0.2$ mags according to the SFD maps, indicating that extinction effects are relatively small for the majority of the LaTE-MoVeRS sample. We consider extinction for a subset of our LaTE-MoVeRS sample in Section~\ref{mirexcess}.

\section{Stars with MIR Excesses}\label{mirexcess}

	As part of an ongoing science campaign to locate field stars (with ages $\gtrsim 1$ Gyr) with large MIR excesses \citep[$L_\mathrm{IR}/L_\ast \gtrsim 10^{-2}$;][]{theissen:2014:146, theissen:2016:2}, we examined the LaTE-MoVeRS sample for stars exhibiting excess 12 \um\ flux (the \emph{WISE} $W3$-band). Some of these stars are believed to have undergone collisions between terrestrial planets or large planetismals \citep{theissen:2014:146}, accounting for the large amounts of excess MIR flux, unlike dusty young objects that may still retain their massive primordial disks. With the proclivity for low-mass stars to host multiple, close-in, terrestrial planets \citep{dressing:2013:95, ballard:2016:66}, collisions are hypothesized to occur due to metastable orbits that become unstable over Gyr timescales \citep{theissen:2014:146, theissen:2016:2}. However, for the lower-mass objects in LaTE-MoVeRS, it is unclear if they host similar quantities of terrestrial planets, although recent observations may indicate similar quantities \citep{gillon:2016:221}. Locating similar large MIR excesses around these lower-mass objects may indicate similar quantities of terrestrial planets (with similar metastable orbits). Furthermore, locating smaller MIR excesses may indicate ultracool, unresolved companions, which have similar sizes, but peak at redder wavelengths.
	
	To cultivate a clean, high-fidelity sample of stars to investigate for MIR excesses, we selected stars that met the following criteria:
	
\begin{enumerate}

\item Stars that did not have a \emph{WISE} extended source flag (\textsc{ext\_flg} = 0), ensuring a point-source morphology through all \emph{WISE} bands (46,226 stars).

\item Stars that did not have a contamination or confusion flag in either $W1$, $W2$, or $W3$ (\textsc{cc\_flg}$_{W1,W2,W3}$ = 0), ensuring clean photometry for those bands (44,397 stars).

\item Stars with at least a signal-to-noise (S/N) of 3 in $W1$, $W2$, and $W3$ (\textsc{WxSNR}$_{\mathrm{x}=W1,W2,W3} \geqslant 3$), ensuring high-probability detections (2,069 stars).

\end{enumerate} 
	
	Next, we visually inspected each object in the \emph{WISE} image archives, removing objects with significant contamination or no apparent $W3$ source. We assigned a quality flag to each true or marginal $W3$ detection, similar to \citet{theissen:2016:2}. We were left with 259 stars showing a good detection (\textsc{quality} = 1) and 963 stars showing a marginal detection (\textsc{quality} = 2).
	
	Extinction effects may be important for this subset of stars when we consider reddening effects due to circumstellar material in addition to interstellar dust. These reddening effects could severally underestimate the flux produced at shorter wavelengths, thereby changing our SED fits. We do not have a way to directly measure the extinction along each line-of-sight between us the our observed stars, therefore, we must use indirect methods to estimate $A_\lambda$. We chose to use the SDSS provided estimates for $A_\lambda$ using the SFD dust maps. This gives us a good approximation of the expected extinction along our line of sight. 

	SDSS bandpass extinction values can be converted into the more commonly used $A_V$ values. We then used the estimated $A_V$ values from the SFD maps to estimate $A_\lambda$ (the extinction per bandpass) using values from the Asiago Database \citep{moro:2000:361, fiorucci:2003:781} for SDSS, 2MASS, and \emph{WISE} bandpasses. Assuming an $R_V = 3.1$, we applied extinction corrections to the $izJHK_SW1W2W3W4$-bands. Further details are provided for our methods in applying extinction corrections to SDSS, 2MASS, and \emph{WISE} bandpasses in \citet{theissen:2014:146}. For the remainder of our analysis we refer to the unreddened photometry. We also include bandpass extinction values ($A_\lambda$) and unreddened photometry for the entire LaTE-MoVeRS sample in the online catalog.
	
\begin{figure}
\centering
\includegraphics[width=\linewidth]{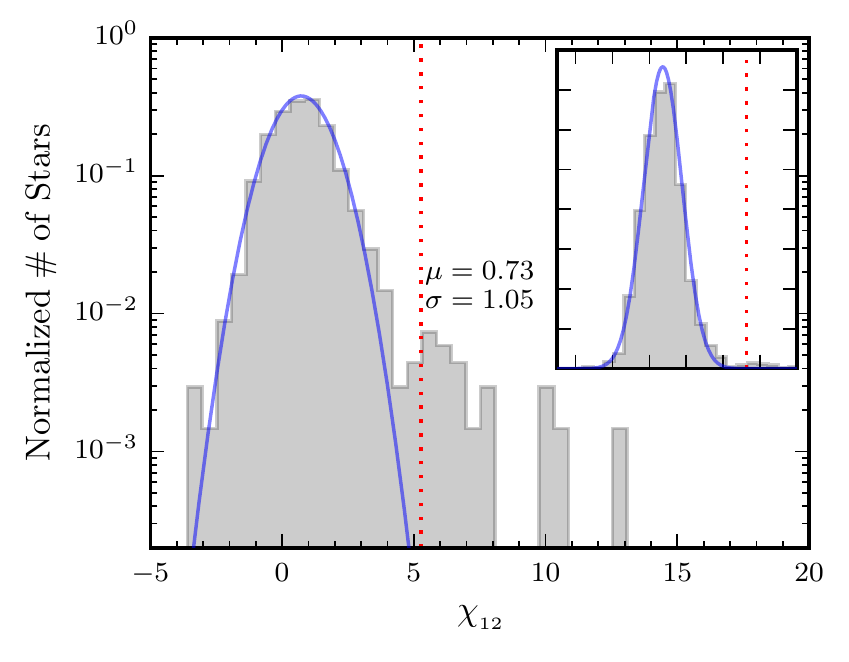}
\caption{Normalized distributions of $\rchi_{12}$. Also plotted is the best-fit normal distribution (blue line). The value for a star considered to have a high-significance MIR excess is shown with the dash-dotted line, $\rchi_{12} > 5.26$ ($4.31\sigma$). The inset plot shows the linear distributions.
\label{fig:chi12}}
\end{figure}

	Numerous studies have used \emph{WISE} photometry to search for and identify stars exhibiting excess MIR flux \citep[e.g.,][]{avenhaus:2012:a105, kennedy:2012:91, patel:2014:10, cotten:2016:15, binks:2016:}. Stars with MIR excesses can be isolated in color-color space when using one color that traces the stellar photosphere (effective temperature/spectral type), and another color that is a combination of the stellar photosphere and the wavelength at which a MIR excess occurs. \citet{avenhaus:2012:a105} developed an empirical relationship based on $V-K_s$ color (a proxy for spectral type) and $W1-W3$, $W1-W4$, and $W3-W4$ colors to determine stars with excess flux at $W3$ and $W4$ (12 and 22 \um) using the RECONS 100 nearest star systems \citep{jao:2005:1954, henry:2006:2360}. This method was shown to be sensitive enough to make a 3$\sigma$ detection of the disk around AU Mic at 22 \um, where other methods using \emph{Spitzer} \citep{plavchan:2009:1068} and \emph{WISE} \citep{simon:2012:114} were unable to make a significant detection at similar wavelengths. This method is also easily applied to extremely large samples of stars, where SED fitting may be computationally intractable. To separate stars exhibiting excess MIR flux, \citet{theissen:2014:146, theissen:2016:2} used an empirical relation based on SDSS $r-z$ color (a proxy for spectral type using SDSS bandpasses) and \emph{WISE} $W1-W3$ color. Due to the lack of $r$-band magnitudes for many sources in the LaTE-MoVeRS sample, no such empirical relationship exists. However, using the stellar parameter estimates from BT-Settl models, we can select for stars with excess MIR flux using the $\rchi_{12}$ metric, defined as
\begin{equation}\label{eqn:excessrelationship}
\rchi_{12} = \frac{F_{12 \mu\mathrm{m,\; measured}} - F_{12 \mu\mathrm{m,\; model}}} {\sigma_{F_{12 \mu\mathrm{m,\; measured}}}}.
\end{equation}	
The distribution of $\rchi_{12}$ values is shown in Figure~\ref{fig:chi12}. Using the above metric, we selected stars with less than 1\% chance of being a false-positive using the equation,
\begin{equation}
P_{FP}(\mathrm{MIR\; Excess}) \times N_\mathrm{sample} < 0.01,
\end{equation}
where $P_{FP}(\mathrm{MIR\; Excess})$ is the probability that the MIR excess is a false-positive, and $N_\mathrm{sample}$ is the number of sources within the sample. This gave us a cutoff values of $\rchi_{12} > 5.26$ ($4.31\sigma$), producing 19 candidate stars exhibiting statistically significant MIR excesses. Six of our MIR excess candidates have $T_\mathrm{eff} \geqslant 2500$ K, and were found in the MIR excess study of \citet{theissen:2016:2}, therefore, we will focus on the 13 MIR excess candidates with $T_\mathrm{eff} < 2500$ K that are not found in the original MoVeRS catalog.

	 We visually inspected the SEDs of the MIR excess candidates compared to their best-fit model photospheres. Six of our candidates appeared to have only marginal MIR excesses when compared to the best-fit model photosphere. A marginal excess could be due to: 1) very small amounts of orbiting dust, similar to AU Mic \citep[e.g.,][]{avenhaus:2012:a105}; 2) a limitation of the BT-Settl models and/or our model parameter space (e.g., not exploring metallicity changes) in reproducing accurate MIR spectra; 3) excesses that span the entire MIR range, possibly indicating either a hotter, closer circumstellar disk than those identified in \citet{theissen:2014:146} or a low-mass binary system \citep{cook:2016:2192}. The SEDs of our remaining seven stars exhibiting MIR excesses are shown in Figure~\ref{fig:seds}. Of these seven objects, only one has not been previously identified as a VLM star or brown dwarf (2MASS J11151597+1937266). Four of these sources also were detected at $W4$. These measurements are consistent with MIR excesses, although they have much larger uncertainties owing to low source counts.

	 To further assess the likelihood that these stars exhibit true excesses in \emph{WISE} bandpasses, we compared the $W1$, $W2$, and $W3$ fluxes to the median empirical SED derived from the 1222 stars in the visually inspected subset. To compute the median empirical SED, we obtained the $izJHK_sW1W2W3$ photometry for all sources with $T_\mathrm{eff}$ estimates within $\pm$200 K of the candidate with a MIR excess, scaled each SED to the $J$-band flux of the candidate object, and computed the median flux for each band. The median empirical SED is shown in Figure~\ref{fig:seds} as the red line. Using this method, all seven objects had $> 3\sigma$ excesses at $W2$ and $W3$, and five objects had $> 3\sigma$ excesses at $W1$. We discuss these objects in the following section.

\begin{figure*}
\centering
\includegraphics{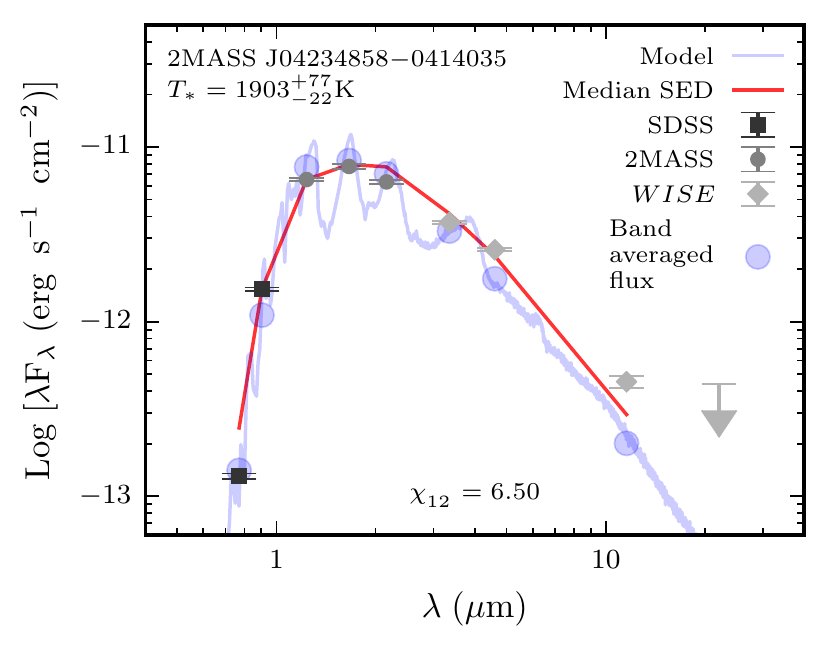}
\includegraphics{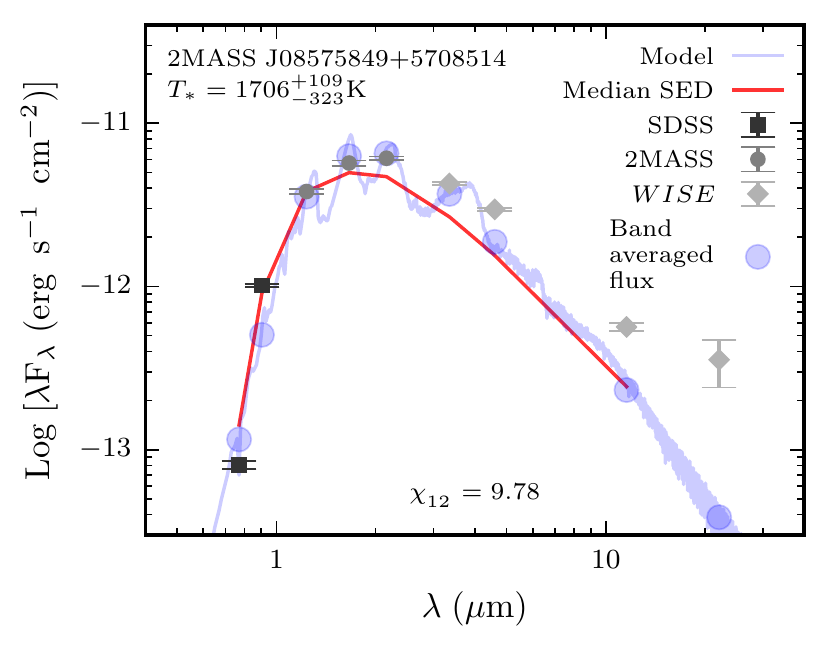}
\includegraphics{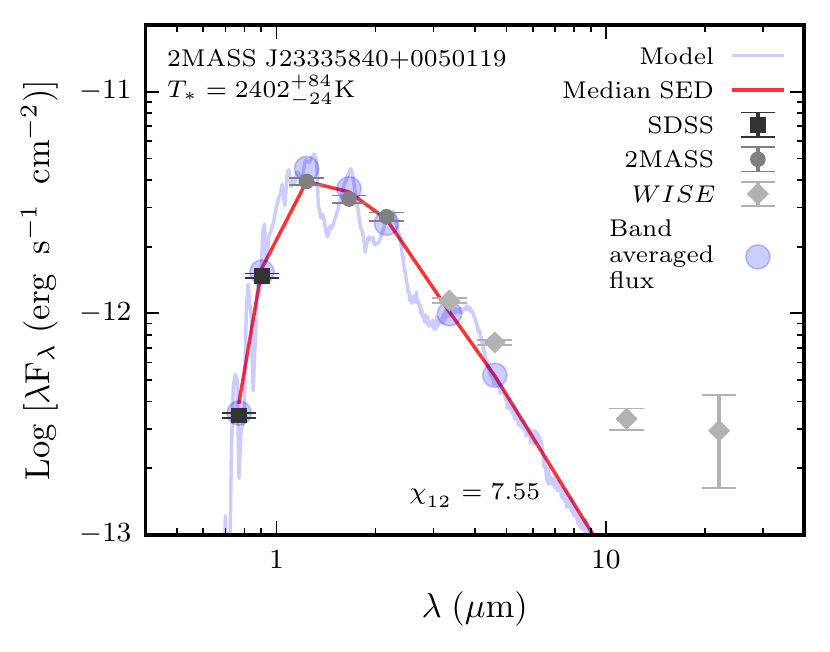}
\includegraphics{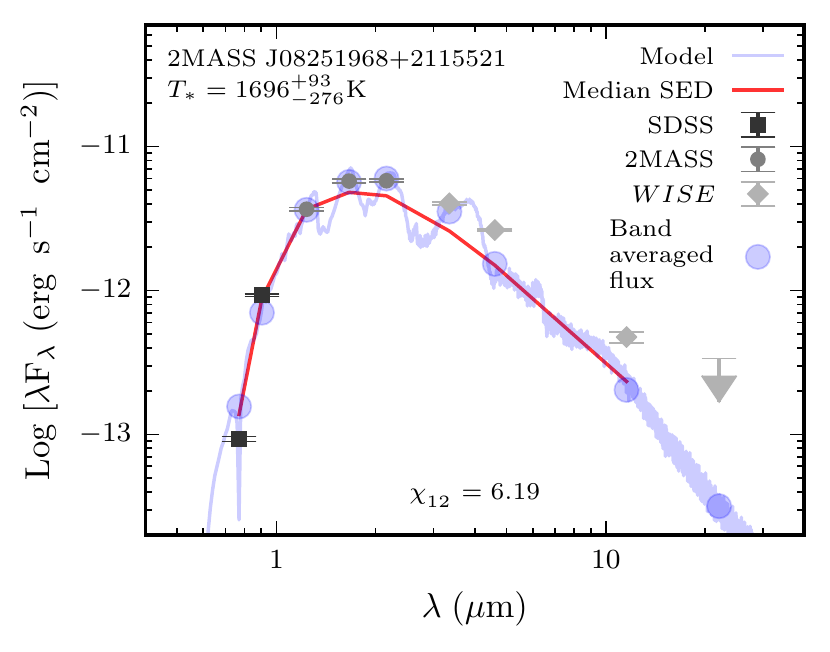}
\captcont{SEDs for the seven previously unidentified objects exhibiting excess MIR flux in our sample. Shown are the best-fit BT-Settl photosphere model (blue line), band-averaged model fluxes (blue circles), median SED (see text for details), and measured fluxes (symbols with error bars). Additionally, we list the MIR excess significance value (defined as $\rchi_{12}$, see text for details). SDSS ObjID and best-fit temperature are listed in the top right.
\label{fig:seds}}
\end{figure*}

\begin{figure*}[!htb]
\centering
\includegraphics{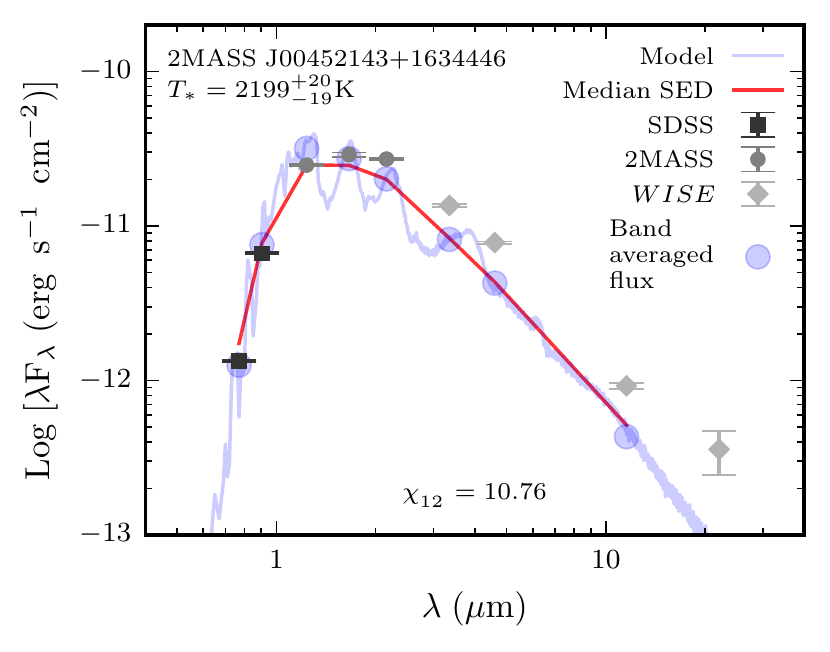}
\includegraphics{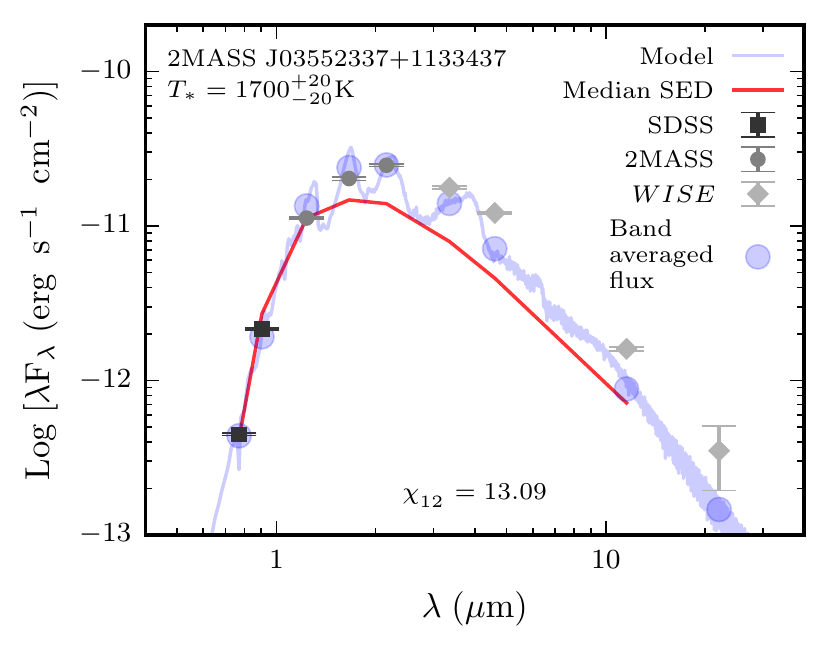}
\includegraphics{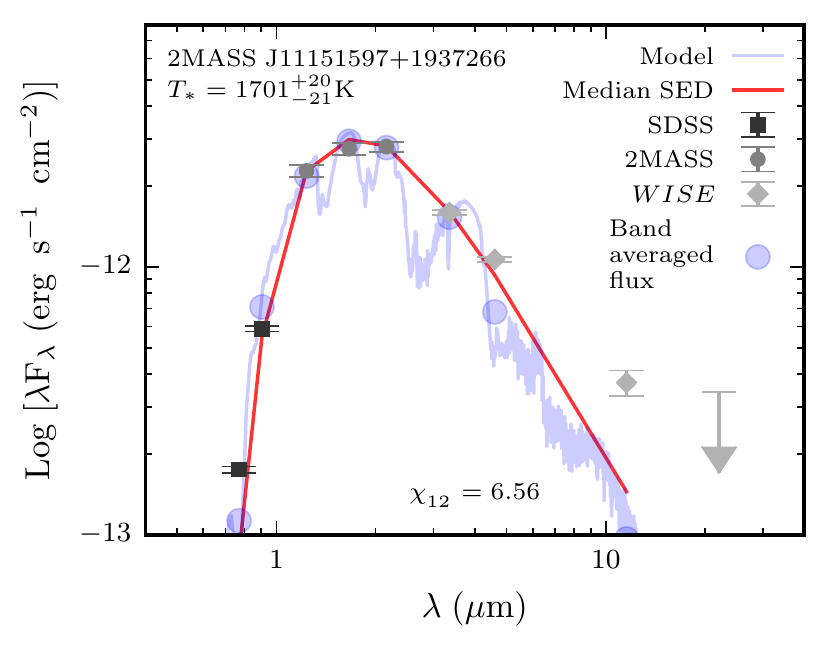}
\caption{continued.}
\end{figure*}

\subsection{Objects with Excess MIR Flux}

\subsubsection{SDSS OBJID 1237649963535958574}

	2MASS J04234858-0414035 was first identified by \citet{geballe:2002:466} as an T0 dwarf. Using \emph{HST} NICMOS, \citet{burgasser:2005:l177} identified this object as a binary system, with a separation of 0.16\arcsec, composed of an L6 primary and T2 secondary. Measuring equivalent widths (EWs), \citet{kirkpatrick:2008:1295} indicated that this binary showed strong H$\alpha$ emission (H$\alpha$ EW $= 3$) and Li \textsc{i} absorption (Li \textsc{i} EW $= 11$), both strong indicators of youth. These measurements were also verified by \citep{pineda:2016:73}. \citet{kao:2016:24} suggested that the low-amplitude $K_s$-band variability of this system \citep{enoch:2003:1006} could be due to auroral activity, which causes localized heating of the atmosphere. Comparing the observed $W3$ flux to the model interpolated $W3$ flux, we find a $W3$ flux ratio ($F_{W3\; \mathrm{(measured)}} / F_{W3\; \mathrm{(model)}}$) of 2.2, and exhibits $> 3\sigma$ excesses at $W2$ and $W3$ when compared to the median SED. This star may represent a way to detect unresolved ultracool binaries using 2MASS and \emph{WISE} photometry, similar to the methods outlined by \citet{cook:2016:2192} for M dwarfs with ultracool companions.

\subsubsection{SDSS OBJID 1237651272963457855}

	2MASS J08575849+5708514 was identified in \citet{geballe:2002:466} as an L8 dwarf. This object was also identified to have a low surface gravity \citep[log $g \approx 4.5$;][]{stephens:2009:154}, although \citet{gagne:2015:33} did not find a clear indication of low surface gravity. However, \citet{gagne:2015:33} posit that the weaker H$_2$($K$) index \citep[defined in][]{canty:2013:2650} for this object versus typical field L8 dwarfs could be an indication of low surface gravity. \citet{gagne:2015:33} did not find a high probability of membership for this object with any NYMG. Furthermore, the kinematics of this extremely red object are more consistent with a field object than any young brown dwarf in the solar neighborhood \citep{gagne:2015:33}. We find a $W3$ flux ratio of 2.6, however, the shape of the $H$-band ``bump" from \citet{gagne:2015:33} is not consistent with the shape expected for an unresolved, ultracool binary \citep[e.g.,][]{bardalez-gagliuffi:2014:143}. We also find $> 3\sigma$ excesses at $W1$, $W2$, and $W3$ when compared to the median SED. Furthermore, the lack of a clear low surface gravity signature makes it unlikely this is a young, dusty brown dwarf \citep[e.g.,][]{faherty:2013:2}, which typically have unusually red NIR colors. The $W4$ detection is consistent with excess MIR flux as compared to the photosphere model, and the $J-K \approx 2.08$ make this L8 dwarf unusually red as compared to other L8 dwarfs \citep[typical $J-K \approx 1.8$;][]{schmidt:2010:1045}. The peculiar properties of this object motivate a deeper investigation.

\subsubsection{SDSS OBJID 1237663462608470296}

	2MASS J23335840+0050119 was previously identified in \citet{theissen:2014:146} for harboring a large MIR excess. This object is a known, magnetically active L0 dwarf \citep{schmidt:2010:1808, west:2011:97, schmidt:2015:158}. \citet{theissen:2014:146} identified this object as having a potential MIR excess, although the image quality was suspect. Further investigation of this object within the \emph{Spitzer} Heritage Archive\footnote{\url{http://irsa.ipac.caltech.edu/applications/Spitzer/SHA}} images reveals an extended object at 3.6 and 4.5 \um, indicating that this object is a blend between a VLM object and most likely a background galaxy not resolved by \emph{WISE}. The excess MIR flux most likely originates from the background source rather than the L0 dwarf.

\subsubsection{SDSS OBJID 1237664835925836294}

	2MASS J08251968+2115521 was identified as an optical L7.5 dwarf by \citet{kirkpatrick:2000:447} and a NIR L6 by \citet{knapp:2004:3553}. \citet{gagne:2015:33} identified this object as a ``peculiar" L7 dwarf with no clear sign of low surface gravity, indicating a potential field object. \citet{bannister:2007:l24} suggest that this object has kinematics consistent with the $\sim$625 Myr Hyades association. We find a $W3$ flux ratio of 2.3, however, due to the presumed age of this object, it is unlikely that the detected MIR excess is due to this being a dusty, young brown dwarf. It is also unlikely that the detected MIR excess is due to a debris disk if fractional IR luminosities of debris disks are similar for brown dwarfs as for higher-mass objects ($L_\mathrm{IR} / L_\ast \approx 10^{-4}$), implying a much smaller $W3$ flux ratio for a debris disk. The MIR excess is supported by comparison to the median SED, which yields $> 3\sigma$ excesses at $W1$, $W2$, and $W3$. The $J-K \approx 2.07$ is unusually red for dwarfs of this spectral type, similar to 2MASS J08575849+5708514, potentially indicating the same mechanism for both these extremely red dwarfs.

\subsubsection{SDSS OBJID 1237679167690375412}

	2MASS J00452143+1634446 is a known low surface gravity L2 dwarf \citep{gagne:2014:121, zapatero-osorio:2014:a6}. This object is a high-probability member of the Argus association \citep{gagne:2014:121}, with an estimated age between 30--50 Myr. \citet{gagne:2014:121} found this object to have unusually red NIR colors, not uncommon for young, late-type dwarfs \citep{faherty:2013:2}. This object has a $W3$ flux ratio of 2, making it comparable to other unusually red, young, ultracool dwarfs, which also accounts for the small $W4$ excess. Additionally, this object has $> 3\sigma$ excesses at $W1$, $W2$, and $W3$ when compared to the median SED.

\subsubsection{SDSS OBJID 1237673327076901196}

	2MASS J03552337+1133437 is a known L4 member of the AB Doradus Moving Group \citep{faherty:2013:2, liu:2013:85, aller:2016:120}. \citet{faherty:2013:2} identified this object as a young, dusty planetary mass object. The observed $W3$ flux ratio of 2.1 is consistent with low surface gravity L dwarfs consistently found as unusually red and underluminous, as compared to their field counterparts. The $W4$ detection also implies an excess of MIR flux, although the large uncertainty makes it a marginal detection. Comparison to the median SED yields $> 3\sigma$ excesses at $W1$, $W2$, and $W3$.

\subsubsection{SDSS OBJID 1237667915950588237}

\begin{figure*}[!htb]
\centering
\includegraphics{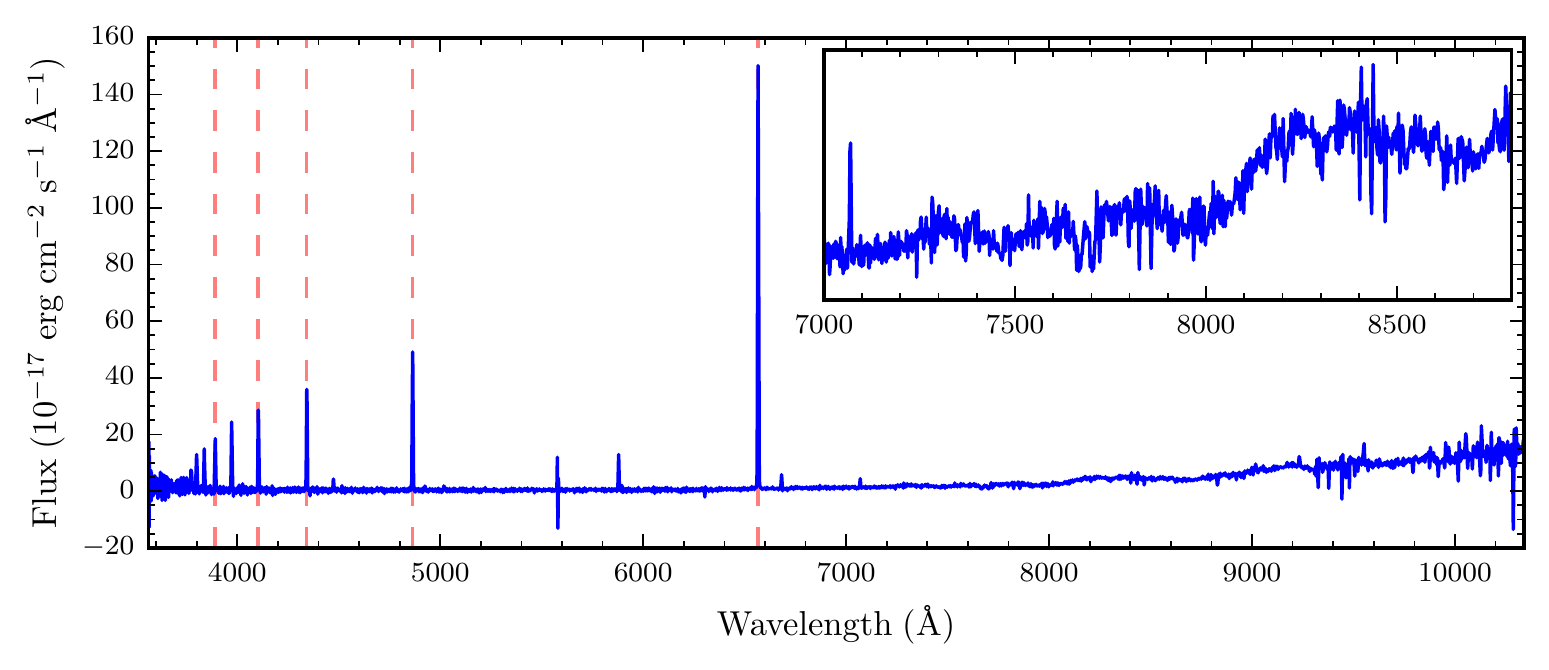}
\caption{SDSS optical spectrum of 2MASS J11151597+1937266. The rest wavelengths for the first few Balmer transitions are indicated with the red dashed lines. The inset plot shows a close up of the red end of the spectrum. The upturn in flux at redder wavelengths indicates this object is very red, consistent with our SED fit.
\label{fig:cv}}
\end{figure*}

	2MASS J11151597+1937266 is not listed in ADS or SIMBAD, however, there is an optical spectrum taken with the SDSS Baryon Oscillation Spectroscopic Survey \citep[BOSS;][]{dawson:2013:10} spectrograph \citep{smee:2013:32}. This object shows a number of strong emission features in its spectrum, typically associated with a flaring event \citep[e.g., H$\alpha$, H$\beta$, H$\gamma$;][]{liebert:1999:345}, as is shown in Figure~\ref{fig:cv}. Additionally, the full width at half maximum of the H$\alpha$ emission line ($\sim$4 \AA) is similar to the width of H$\alpha$ emission lines from field dwarfs \citep{west:2011:97}, indicating no broadening due to accretion.
	
\begin{figure*}[!htb]
\centering
\includegraphics[width=1\textwidth]{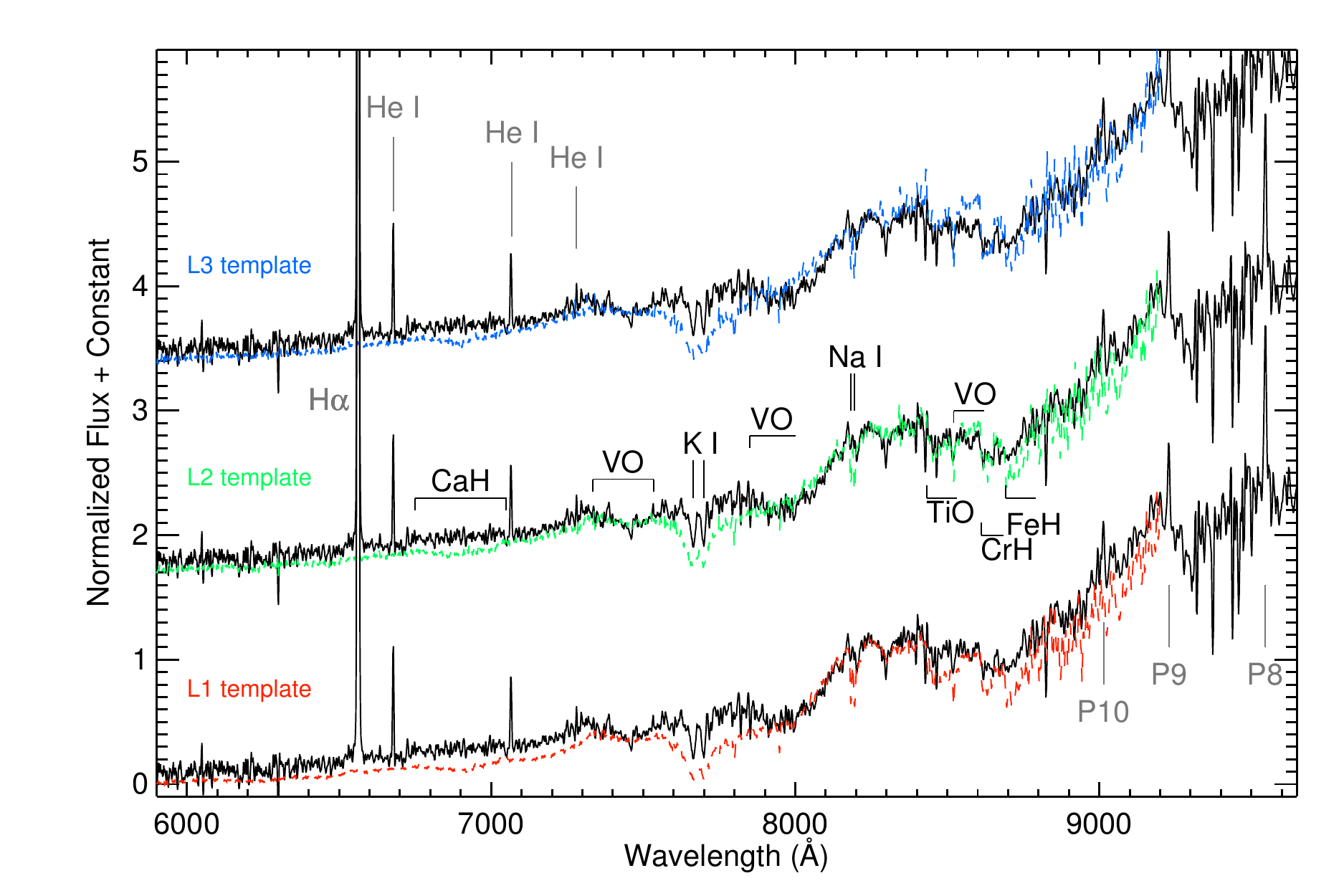}
\caption{Comparison of 2MASS J11151597+1937266 (black) to L dwarf template spectra (gray) from \citet{schmidt:2014:642}. The best-fit spectral type is the L2 template. Photospheric line identifications in black are from \protect\citet{kirkpatrick:1999:802}. Line identifications most likely associated with a flaring event are indicated in gray \protect\citep{liebert:1999:345}. 
\label{fig:l2gamma}}
\end{figure*}

\begin{figure*}[!htb]
\centering
\includegraphics[width=1\textwidth]{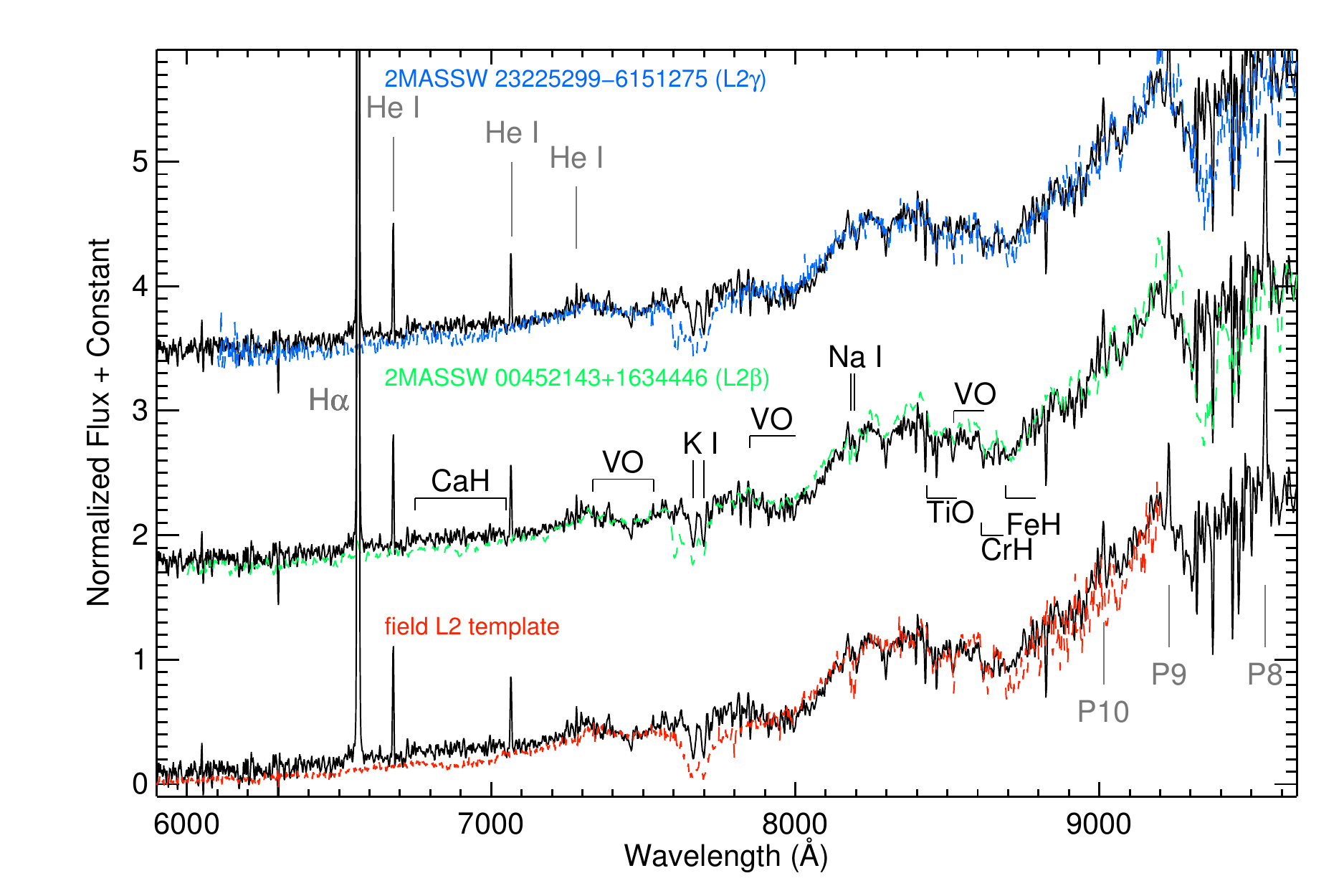}
\caption{Similar to Figure~\ref{fig:l2gamma}. Comparison of 2MASS J11151597+1937266 (black) to L dwarf spectra with different surface gravities \protect\citep{cruz:2009:3345}, starting from the highest surface gravity (bottom) to the lowest surface gravity (top). Comparison spectra are from \protect\citet{reid:2008:1290} and \protect\citet{schmidt:2014:642}. The best-fit spectral type is an L2$\gamma$, however, none of the spectra adequately fit the K \textsc{i} feature. This may indicate that the object is lower gravity than the L2$\gamma$ standard, or that there is some veiling in the continuum.
\label{fig:l2gamma2}}
\end{figure*}

	Comparing to field-age L dwarf templates \citep{schmidt:2014:642} and young L dwarf standards \citet{reid:2008:1290}, we found the best-fit spectral type to be an L2$\gamma$ \citep{cruz:2009:3345}, as is shown in Figure~\ref{fig:l2gamma} and Figure~\ref{fig:l2gamma2}, consistent with this object being young (low-surface gravity). Despite the good overall match between the object and the L2$\gamma$, especially around 8500\AA, the K \textsc{i} lines ($\sim$7700\AA) of the object appear to be significantly weaker. This could indicate that the object is lower gravity than the L2$\gamma$ standard, but may instead signal that there is some veiling in the continuum (as indicated by the elevation of the continuum above the template spectrum between 6000--8000\AA). Our measured $W3$ flux ratio of 3.7 may also indicate the presence of circumstellar material around this object.
	
	To determine if this star belongs to a known NYMG, we computed its 3-dimensional space velocities ($UVW$). We measured the radial velocity using a cross-correlation-like program \citep[\emph{xcorl.py};][]{mohanty:2003:451, west:2009:1283, theissen:2014:146}, comparing to the optical spectrum of 2MASS J23225299$-$6151275 \citet{cruz:2009:3345}. Our measured radial velocity and uncertainty are listed in Table~\ref{tbl:candidate}, along with all the kinematic information for 2MASS J11151597+1937266. We used the BANYAN II webtool \citep{malo:2013:88, gagne:2014:121} to determine the probability of this object being associated with a known kinematic group. The kinematics of this object do not make it a likely member of any of the known kinematic associations contained in the BANYAN II webtool (probabilities of association $\ll 1$). However, according to the BANYAN II webtool, the kinematics of this object are also not consistent with a field object (young or old). We do not rule out the possibility that this is a field object, although our spectroscopic analysis indicates that this is a young object. The kinematics of this object may be explained through dynamical ejection from a young association \citep[e.g.,][]{reipurth:2001:432}.
	
\begin{deluxetable}{cc}
\tabletypesize{\footnotesize}
\tablecolumns{2}
\tablecaption{Properties of 2MASS J11151597+1937266\label{tbl:candidate}}
\tablehead{
\colhead{Parameter} & \colhead{Value} 
}
\startdata
SDSS DR8$+$ objID		& 1237667915950588237\\
R.A.					& $168.816447^\circ$\\
Dec.					& $19.624012^\circ$\\
$T_\mathrm{eff}$		& $1701^{+20}_{-21}$ K\\
Distance				& $48 \pm 6$ pc\\
Radial Velocity			& $27 \pm 7$ km s$^{-1}$\\ 
$\mu_\alpha$ ($\cos \delta$)			& $-57 \pm 13$ mas yr$^{-1}$\\
$\mu_\delta$ 			& $-25 \pm 8$ mas yr$^{-1}$\\ 
$U$\tablenotemark{a}	& $-5 \pm 4$ km s$^{-1}$\\
$V$					& $-5 \pm 3$ km s$^{-1}$\\
$W$					& $-26 \pm 7$ km s$^{-1}$\\
$\frac{F_{W3\; \mathrm{(measured)}}}{F_{W3\; \mathrm{(model)}}}$	&	3.7
\enddata
\tablenotetext{a}{Positive values indicate motion towards the Galactic center, putting the $UVW$ space motions into a right-handed coordinate system.}
\end{deluxetable}

\section{Summary and Discussion}\label{discussion}

	We have built a catalog of \latemovers\ late-type ($>$ M5) objects with verified proper motions, termed the Late-Type Extension to the Motion Verified Red Stars (LaTE-MoVeRS) catalog. Proper motions were computed using the SDSS, 2MASS, and \emph{WISE} datasets following similar methods to those outlined in \citet{theissen:2016:41}. We required all of the stars within the LaTE-MoVeRS sample to have significant proper motions ($\mu_\mathrm{tot} \geqslant 2\sigma_{\mu_\mathrm{tot}}$). The LaTE-MoVeRS sample has a typical proper motion precision of $\sim$13 mas yr$^{-1}$. Comparison of the LaTE-MoVeRS proper motions to other proper motion catalogs (MoVeRS, BASS, LSPM, PPMXL, and URAT1) showed extremely good agreement between both proper motion components. Proper motions for cross-matched stars typically had more than 95\% agreement at the 2$\sigma$ level.
	
	Using \emph{Gaia} DR1, we cross-matched the LaTE-MoVeRS sample to the 1,142,679,769 objects within DR1. We obtained 29,535 matches and found that \emph{Gaia} is missing $\sim$68\% of the VLM stars and brown dwarfs with $i \leqslant 20$ within the combined SDSS+2MASS+\emph{WISE} dataset. Additionally, the fraction of matches to \emph{Gaia} falls below 30\% for sources with $i > 20$, independent of color (spectral type).
	
	The LaTE-MoVeRS sample is primarily made of late-type M dwarfs (M6--M8), but extends to late-L spectral types. Using photometric parallax relationships from S. J. Schmidt et al. (2016, in preparation), we computed distances for all the objects in the LaTE-MoVeRS sample. The median distance for objects within the sample is 153 pc, with the closest and farthest objects estimated at 7.8 pc (excluding the potential AGB star with an estimated distance of $\sim$1 pc) and 418 pc, respectively. We also identified 13 new systems with estimated distances $\leqslant$ 25 pc. We estimated $T_\mathrm{eff}$ values for the entire LaTE-MoVeRS sample using BT-Settl models and the \emph{emcee} Python package and the methods outlined in \citet{theissen:2016:2}. The objects in the LaTE-MoVeRS sample typically have $1600 \lesssim T_\mathrm{eff} \lesssim 3800$ K, compared to the original MoVeRS sample which covers $2500 \lesssim T_\mathrm{eff} \lesssim 4800$ K. 
	
	As part of an ongoing effort to search for field objects exhibiting large amounts of excess MIR flux, we selected one newly identified object with $T_\mathrm{eff} < 2500$ K exhibiting excess MIR flux above the expected photospheric value. This object appears to be a young, active, low-surface gravity L2 dwarf exhibiting a large amount of excess MIR flux. If this object is very young ($\lesssim 10$ Myr), the detected MIR excess could be due to a primordial disk. The kinematics of this object are inconsistent with any known NYMG, indicating a possible field object. This dwarf may represent a dynamically ejected object from a young association.
	
	The objects contained within the LaTE-MoVeRS sample represent the bottom of the main sequence and beyond. Due to the limitations of \emph{Gaia}, the LaTE-MoVeRS sample is invaluable for kinematic and nearby studies of VLM stars and brown dwarfs. The LaTE-MoVeRS catalog is available through SDSS CasJobs and VizieR.

\acknowledgments

	The authors would like to thank the anonymous referee for their extremely helpful comments and suggestions. The authors would also like to thank Jonathan Gagn\'e, Daniella Bardalez Gagliuffi, Julie Skinner, Dylan Morgan, and Christian Aganze for their helpful discussions. C.A.T. would like to acknowledge the Ford Foundation for his financial support. A.A.W acknowledges funding from NSF grants AST-1109273 and AST-1255568. A.A.W. and C.A.T. also acknowledges the support of the Research Corporation for Science Advancement's Cottrell Scholarship. This material is based upon work supported by the National Aeronautics and Space Administration under Grant No. NNX16AF47G issued through the Astrophysics Data Analysis Program.

	Funding for the Sloan Digital Sky Survey IV has been provided by
the Alfred P. Sloan Foundation, the U.S. Department of Energy Office of
Science, and the Participating Institutions. SDSS-IV acknowledges
support and resources from the Center for High-Performance Computing at
the University of Utah. The SDSS web site is www.sdss.org.

SDSS-IV is managed by the Astrophysical Research Consortium for the 
Participating Institutions of the SDSS Collaboration including the 
Brazilian Participation Group, the Carnegie Institution for Science, 
Carnegie Mellon University, the Chilean Participation Group, the French Participation Group, Harvard-Smithsonian Center for Astrophysics, 
Instituto de Astrof\'isica de Canarias, The Johns Hopkins University, 
Kavli Institute for the Physics and Mathematics of the Universe (IPMU) / 
University of Tokyo, Lawrence Berkeley National Laboratory, 
Leibniz Institut f\"ur Astrophysik Potsdam (AIP), 
Max-Planck-Institut f\"ur Astronomie (MPIA Heidelberg), 
Max-Planck-Institut f\"ur Astrophysik (MPA Garching), 
Max-Planck-Institut f\"ur Extraterrestrische Physik (MPE), 
National Astronomical Observatory of China, New Mexico State University, 
New York University, University of Notre Dame, 
Observat\'ario Nacional / MCTI, The Ohio State University, 
Pennsylvania State University, Shanghai Astronomical Observatory, 
United Kingdom Participation Group,
Universidad Nacional Aut\'onoma de M\'exico, University of Arizona, 
University of Colorado Boulder, University of Oxford, University of Portsmouth, 
University of Utah, University of Virginia, University of Washington, University of Wisconsin, 
Vanderbilt University, and Yale University.
 
	This publication makes use of data products from the Two Micron All Sky Survey, which is a joint project of the University of Massachusetts and the Infrared Processing and Analysis Center/California Institute of Technology, funded by the National Aeronautics and Space Administration and the National Science Foundation. This publication also makes use of data products from the \emph{Wide-field Infrared Survey Explorer}, which is a joint project of the University of California, Los Angeles, and the Jet Propulsion Laboratory/California Institute of Technology, funded by the National Aeronautics and Space Administration. 
 
	This work has made use of data from the European Space Agency (ESA)
mission {\it Gaia} (\url{http://www.cosmos.esa.int/gaia}), processed by
the {\it Gaia} Data Processing and Analysis Consortium (DPAC,
\url{http://www.cosmos.esa.int/web/gaia/dpac/consortium}). Funding
for the DPAC has been provided by national institutions, in particular
the institutions participating in the {\it Gaia} Multilateral Agreement.

	This research has benefitted from the Ultracool RIZzo Spectral Library \citep[\url{http://dx.doi.org/10.5281/zenodo.11313};][]{cruz:2014:}, maintained by Jonathan Gagn\'e and Kelle Cruz.
  
	The authors are also pleased to acknowledge that much of the computational work reported on in this paper was performed on the Shared Computing Cluster which is administered by Boston University's Research Computing Services (\url{www.bu.edu/tech/support/research/}). This research made use of Astropy, a community-developed core Python package for Astronomy \citep{astropy-collaboration:2013:a33}. Plots in this publication were made using Matplotlib \citep{hunter:2007:90}. This research has made use of the SIMBAD database, operated at CDS, Strasbourg, France \citep{wenger:2000:9}. This research has also made use of the VizieR catalogue access tool, CDS, Strasbourg, France. This research has also made use of NASA's Astrophysics Data System.

\bibliography{ms}
\bibliographystyle{apj}

\end{document}